\documentclass[11pt,a4paper]{article}
\pdfoutput=1
\usepackage{jheppub}
\usepackage{gensymb}
\usepackage{subcaption}
\usepackage{mwe}
\usepackage{amssymb,amsmath}
\usepackage{graphicx}
\usepackage{color}
\usepackage{cancel}
\usepackage{booktabs}
\usepackage[colorlinks=true
,urlcolor=blue
,citecolor=blue
,linkcolor=blue
,pagecolor=blue
,linktocpage=true
,pdfproducer=medialab
]{hyperref}

\usepackage[numbers]{natbib}
\usepackage{notoccite}
\usepackage{comment} 
\usepackage{placeins}
\usepackage{multirow}
\usepackage{setspace}
\usepackage{bigints}
\makeatletter \renewcommand{\@dotsep}{10000} \makeatother
\newcommand{\PS}{SU(4)_{C}\times SU(2)_{L}\times SU(2)_{R}}
\newcommand{\mgut}{M_{{\rm GUT}}}
\newcommand{\msusy}{M_{{\rm SUSY}}}

\newcommand{\FTT}{4-2-2}
\newcommand{\FT}{\Delta_{{\rm EW}}}
\newcommand{\cred}[1]{{\bf \color{red} #1}}

\DeclareUnicodeCharacter{2212}{-}
\DeclareUnicodeCharacter{202F}{\,}
\begin{document}

\begin{titlepage}
\pagestyle{empty}

\vspace*{0.2in}
\begin{center}
{\Large \bf Compressed Stop-Neutralino Spectra from \\ Yukawa Unified Non-Holomorphic Pati-Salam Model: \\ Prospects for the FCC}

\vspace{1cm}
{\bf Ali \c{C}i\c{c}i\footnote{E-mail: ali.cici@cern.ch}$^{,a,b}$, B\"{u}\c{s}ra Ni\c{s}\footnote{E-mail:busranis@uludag.edu.tr}$^{,c}$ and Cem Salih $\ddot{\rm U}$n\footnote{E-mail: cemsalihun@uludag.edu.tr}$^{,c}$}
\vspace{0.5cm}

{\small \it  $^{a}$Orhaneli T\"{u}rkan-Sait Y\i lmaz High School, TR16980 Bursa,T$\ddot{u}$rkiye 
\\  $^b$Department of Physics, Faculty of Engineering and Natural Sciences, \\ Bursa Technical University, TR16310, Bursa, T\"{u}rkiye
\\ $^{c}$Department of Physics, Bursa Uluda\~{g} University, TR16059 Bursa, T$\ddot{u}$rkiye}

\end{center}

\vspace{0.5cm}
\begin{abstract}

We explore the weak-scale phenomenology of supersymmetric models based on the $SU(4)_{C}\times SU(2)_{L}\times SU(2)_{R}$ gauge symmetry. We include non-holomorphic soft supersymmetry-breaking terms arising from perturbations on D-branes. These terms significantly alter the implications of Yukawa unification, as they directly interfere in the threshold corrections to the Yukawa couplings. With these corrections, Yukawa unification can be compatible with low fine-tuning alongside a heavy Higgsino-like lightest supersymmetric particle; however, these solutions are strongly constrained by dark matter observations. Furthermore, in contrast to previous studies, the non-holomorphic contributions accommodate heavy gluino masses in this class of models from approximately 2.2 to 10 TeV while preserving Yukawa unification. These gluinos can be probed up to about 2.5 TeV in high-luminosity collider searches, and up to about 6 TeV at future 100 TeV center-of-mass energy colliders. With the non-holomorphic threshold corrections to Yukawa couplings, Yukawa unification can be accommodated with relatively light third-generation squarks. We find that the supersymmetric spectra can accommodate sbottom masses around 1.5 TeV, making them highly accessible to upcoming experimental searches. The stop can also be as light as about 1.5~TeV, resulting in a compressed stop-neutralino spectrum. Even though these solutions lie beyond the sensitivity of current collider searches, they can be potentially probed at future facilities, such as the proposed Future Circular Collider. Our results do not include systematic uncertainties, which can heavily impact the experimental analysis. In this context, our findings serve to highlight potential directions and prospects for new physics searches at future collider experiments. Our results assume that overall systematic uncertainties in background modeling do not exceed $0.1\%$.

\end{abstract}
\end{titlepage}

\section{Introduction}
\label{sec:intro}

Despite excessive studies on supersymmetry (SUSY) and the absence of direct signals from the ongoing experiments, it is still one of the appealing candidates for new physics models beyond the Standard Model (BSM). SUSY models provide dynamically interfering new particles in experimentally testable processes such as pair productions in collider experiments \cite{ATLAS:2023afl,CMS:2021eha}, rare decays of $B-$meson \cite{Boubaa:2022xsk}, muon anomalous magnetic moment (muon $g-2$) \cite{Moroi:1995yh,Martin:2001st,Giudice:2012pf} etc. From the theoretical point of view, one of the main motivations behind SUSY is the solution to the hierarchy problem, and if one assumes soft SUSY breaking (SSB), the supersymmetric models can still provide a cancellation in quadratic divergent contributions to the Higgs boson mass \cite{Gildener:1976ai,Gildener:1979dd,Weinberg:1978ym,Susskind:1978ms,Veltman:1980mj,Susskind:1982mw,Dvali:1995qt}. Besides, the 125 GeV Higgs boson mass destabilizes the vacuum at about $10^{10} $ GeV \cite{Elias-Miro:2011sqh} within the Standard Model (SM), while the stability of the Higgs potential can be ensured in SUSY models \cite{Degrassi:2012ry,Bezrukov:2012sa,Buttazzo:2013uya,Kitahara:2013lfa,Carena:2012mw}. In addition to such strong motivations, the gauge couplings of the Standard Model (SM) can unify at a grand unification scale ($\mgut$). In this context, one can construct supersymmetric grand unified theories (SUSY GUTs) and explore possible high scale origins of the particle phenomena observed or expected at the low-scale experiments.

Among SUSY GUTs, those based on $SO(10)$ provide a particularly interesting scheme of unification. In addition to the gauge couplings, all the matter fields of a generation can be unified into a single 16-dimensional spinor representation. Unifying all the fields in a single representation also allows to unify the Yukawa interactions in its minimal construction (hereafter refers to Yukawa Unification - YU) \cite{Ananthanarayan:1991xp,Ananthanarayan:1992cd}. The first thing to note is that YU fails to be consistent with the observed masses of the fermions from the first two generations \cite{Langacker:1980js,Babu:1992ia}. This conflict can be resolved by extending the Higgs sector at $\mgut$ by including fields from different $SO(10)$ representations (for a recent reviews see, for instance, \cite{Antusch:2019avd,Saad:2025cfb}). In such cases, the MSSM Higgs doublets, in general, turn out to be a superposition of these fields and hence YU is broken. {However, YU can be maintained for the third generation if one assumes that the MSSM Higgs doublets reside solely within the 10-Dimensional ($10_{H}$) representation of $SO(10)$}, the third generation fermions acquire their masses mostly from the vacuum expectation values (VEVs) in $10_{H}$ \cite{AdeelAjaib:2013dnf,Joshipura:2012sr}. 

Once YU is maintained for the third generation, it yields a significant impact on the low-scale implications. This impact arises from its sensitivity to the threshold corrections to the Yukawa couplings at the scale where SUSY particles decouple from the spectrum ($\msusy \equiv \sqrt{m_{\tilde{t}_{1}}m_{\tilde{t}_{2}}}$). YU requires large negative threshold corrections to $y_{b}$ \cite{Gogoladze:2010fu}, which can directly shape the masses of stops, sbottom, gluinos and Higgsinos as well as $\tan\beta$ and SSB trilinear scalar interaction couplings \cite{Pierce:1996zz}. Even though its impacts on the low-scale implications have been extensively explored over years, they are still of interest in constraining the SUSY models \cite{Antusch:2025rbp,Gao:2025gyo,Hussain:2025bxp,Li:2025ypb,Shafi:2023sqa,Ahmed:2022ibc,Aboubrahim:2021phn}.

Within these studies, the supersymmetric high-scale models based on $\PS$ gauge symmetry (\FTT~ for short) \cite{Pati:1974yy,Lazarides:1980tg,Kibble:1982ae} form an interesting class. Its gauge group is the maximal subgroup of $SO(10)$ and emerges if $SO(10)$ breaks through the VEVs in the $54_{H}$ and/or $210_{H}$ representations \cite{Kibble:1982dd,Lazarides:1985my,Babu:2016bmy}. The models in this class can also be embedded in string theory with intersecting $D-$branes \cite{Mansha:2025gvr,Leontaris:2025yuh,Li:2022cqk,Florakis:2021bws,Antoniadis:1990hb}. The \FTT~ model preserves most of the salient features of $SO(10)$ such as the conservation of $B-L$ symmetry, a kind of matter unification (the matter fields are divided into two representations according to their handedness) and YU in its minimal construction. Even though it does not have to be classified as a grand unified theory (GUT), if its symmetry breaks into the SM gauge group near $\mgut$, the gauge coupling unification can be maintained approximately. If the breaking of \FTT~ happens through VEVs in $126$ and $\overline{126}$ representations, then a $Z_{2}$ symmetry is preserved, which acts as the matter parity such that the lightest supersymmetric particle (LSP) happens to be stable \cite{Kibble:1982dd}. Despite yielding rich phenomena \cite{Gogoladze:2009bn,Shafi:2023ksr,Gomez:2020gav,Shafi:2015lfa,Karagiannakis:2012sv}, the YU condition significantly constrains the available parameter space, because it leads to heavy mass spectra for the supersymmetric scalars \cite{Gogoladze:2011db,Blazek:2003wz,Raza:2014upa}. Therefore, probing YU solutions necessitates much powerful colliders such as those proposed in Future Circular Collider (FCC) \cite{FCC:2025lpp,FCC:2025uan,FCC:2025jtd}. One of our previous analyses has shown that FCC can probe the stops in proton-proton collisions up to about 2.5 TeV at $95\%$ Confidence Level (CL), and about 3 TeV at $68\%$ CL \cite{Altin:2020qmu}. In contrast to the heavy scalar spectrum, YU in the minimal setup for these models yield relatively light gluinos ($m_{\tilde{g}} \lesssim 1.5$ TeV), and most of these solutions are excluded by gluino searches ($m_{\tilde{g}} \gtrsim 1.4$ TeV when it is next to LSP (NLSP) \cite{ATLAS:2021yij,ATLAS:2021twp,ATLAS:2020syg}). 

In our work, we revisit this class of Yukawa unified models by incorporating non-holomorphic (NH) SSB terms. Although, these terms are strictly forbidden by the holomorphy condition, they can emerge after SUSY breaking \cite{Inoue:1982pi,Hall:1990ac,Bagger:1993ji,Bagger:1995ay,Martin:1999hc,Jack:1999ud,Jack:1999fa,Haber:2007dj}. While the presence of these terms can potentially be a sign for the ``hard" SUSY breaking which may reintroduce the hierarchy problem, they can be classified in SSB terms if the messenger fields mediating the SUSY breaking to the visible sector are not singlet under the SM gauge group \cite{Inoue:1982pi,Martin:1999hc,Jack:1999ud,Jack:1999fa}. However, when the NH terms are induced by the SUSY breaking, they are suppressed by the SUSY breaking scale in the visible sector \cite{Bagger:1995ay,Martin:1999hc,Haber:2007dj}; thus, their impact can be negligible in comparison with the holomorphic SSB terms. On the other hand, this suppression does not hold, when the NH terms are induced by the perturbations on $D-$branes \cite{deWit:1996kc,Bellisai:1997ck,Dine:1997nq,Gonzalez-Rey:1998xrq,Buchbinder:1998qd,McGuirk:2012sb} in string theories. 

Recent analyses exploring NH contributions \cite{Chattopadhyay:2017qvh,Ali:2021kxa,Chakraborty:2019wav,Un:2014afa,Rehman:2022ydc,Israr:2025cfd,Israr:2024ubp,Rehman:2025djc,Rehman:2024tdr,Un:2023wws,Hicyilmaz:2021onw,Nis:2025fxc} provide a promising framework for exploring NH impact on YU and its low-scale implications. Even though the renormalization group equations (RGEs) for the Yukawa couplings are not modified by the NH terms directly, YU still remains sensitive to these terms through the threshold corrections. In this context, one can realize entirely different regions for YU in the fundamental parameter space, which lead to drastically different implications for the low-scale observations. In our work, their effects on the SUSY mass spectra can be of special interest for the direct detection and probing the SUSY particles. YU, in general, favors heavy mass spectra for strongly interacting SUSY scalars, which can even be beyond the reach of future colliders. In this case, if the NH terms can reduce their masses considerably, the scenarios compatible with YU can become accessible in collider experiments. In our work, we examine possible collider probes after discussing the phenomenological impacts of the NH terms. 

The rest of the paper is organized as follows: We first discuss the NH threshold corrections to Yukawa couplings which are crucial for YU in Section \ref{sec:NHYU}. We also briefly discuss if the naturalness of SUSY models can be satisfied. Our numerical and collider analyses together with the experimental constraints are summarized in Section \ref{sec:scan}. We start the discussion of the NH contributions by comparing their results with the holomorphic cases, and explore possible probes in Section \ref{sec:FPUYU}. Section \ref{sec:stopcoll} discusses YU compatible mass spectra for the SUSY particles with special emphasis on the stop and LSP neutralino mass scales. After we go into detailed detector analyses to probe stops, we summarize and conclude our findings in Section \ref{sec:conc}.

\section{NH Contributions to the Yukawa Unification}
\label{sec:NHYU}

\begin{figure}[h!]
\centering
\includegraphics[scale=0.7]{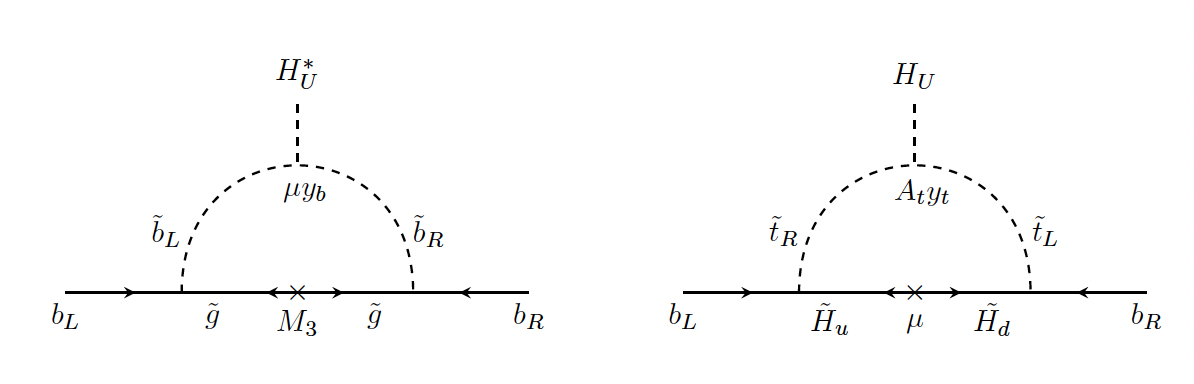}
\caption{The threshold corrections to the b-quark Yukawa coupling.}
\label{fig:ybth}
\end{figure}

One of the first observations about the impact of NH terms is that regions of the parameter space previously excluded by the experimental constraints can become viable again once NH contributions are accounted for \cite{Un:2014afa,Nis:2025fxc}. Even though they can widely open the consistent parameter space, we consider the more direct NH contributions to the Yukawa couplings that accommodate YU solutions within models in \FTT~ class. As mentioned before, realizing YU solutions requires large negative threshold corrections to $y_{b}$, which are diagrammatically represented in Figure \ref{fig:ybth}.

The parameter $M_{3}$ in the left diagram denotes the SSB mass term for the gluino as $m_{\tilde{g}} = |M_{3}|$. The $\mu-$term is involved in the left diagram from the mixing of the SUSY scalars, and from the Higgsino mass in the right diagram. Note that the diagrams display only the holomorphic contributions, which can be calculated as \cite{Hall:1993gn}:

\begin{equation}
\delta_{y_{b}}^{H} \approx \dfrac{g_{3}^{2}}{12\pi^{2}}\dfrac{\mu M_{3}\tan\beta}{m_{\tilde{b}}^{2}}+\dfrac{y_{t}^{2}}{32\pi^{2}}\dfrac{\mu A_{t}\tan\beta}{m_{\tilde{t}}^{2}}~~,
\label{eq:deltab}
\end{equation}
where $m_{\tilde{b}}^{2} = m_{\tilde{b}_{L}}m_{\tilde{b}_{R}}$ and $m_{\tilde{t}}^{2} = m_{\tilde{t}_{L}}m_{\tilde{t}_{R}}$. The first term provides considerable positive contributions when $\mu$ and $M_{3}$ are set to be positive in numerical calculations. Consequently, it should be suppressed to realize the YU solutions, and such a suppression can be realized with light gluinos ($M_{3} \lesssim 1.2$ TeV) and heavy sbottoms ($m_{\tilde{b}} \gtrsim 5$ TeV) \cite{Gogoladze:2009ug}. One can also consider lower values for $\mu$ to minimize the positive contributions, but this parameter also enters the second term which can account for the desired negative contributions. Since the second term is the only source of the negative contributions, YU restricts the trilinear SSB scalar interaction term ($A_{t}$) to be negative. However, if one sets $\mu$ to low values, then the required negative threshold corrections necessitate so large $A_{t}$ values that lead to color/charge breaking (CCB) vacua \cite{Ellwanger:1999bv,Camargo-Molina:2013sta}. Indeed, previous studies have shown that YU favors large $\mu$ values along with negative $A_{t}$, and the positive contributions from the first term are suppressed by light gluino and heavy sbottom masses.

The impact of incorporating the NH terms into the threshold corrections can be understood from the following NH SSB Lagrangian \cite{Jack:1999ud,Jack:1999fa}:

\begin{equation}
\mathcal{L}^{{\rm NH}}_{{\rm soft}} =  -\mu^\prime {\tilde H_u}\cdot
{\tilde H_d} -T_{u}^{\prime}\tilde{Q}~{H}_d^{\dagger} \tilde{U}-
T^\prime_{d}\tilde{Q}~ {H}_u^{\dagger}  \tilde{D}
 - T^\prime_{e}\tilde{L}~{H}_u^{\dagger} \tilde{E} - \mbox{h.c.}~,
\label{eq:SSBLag}
\end{equation}

{where $\mu^{\prime}$ is the NH term contributing to the Higgsino mass, while $T_{i}^{\prime}$ are similar to the holomorphic trilinear scalar coupling while they couple the supersymmetric matter fields to the ``wrong" Higgs field. Even though these terms are strictly forbidden in supersymmetric models, they can be induced from higher dimensional operators through SUSY breaking \cite{Bagger:1995ay,Martin:1999hc,Haber:2007dj}; however, these operators lead to a strong suppression for the NH terms by the SUSY breaking scale at the visible sector. In our work, we consider another approach in which the NH terms arise from the non-supersymmetric perturbations in the string theory \cite{Becker:1996gj,Dasgupta:1999ss,Greene:2000gh,Giddings:2001yu,McGuirk:2012sb}. If one follows the Giddins-Kachru-Polchinski (GKP) compactification \cite{Giddings:2001yu}, the hierarchy between the SUSY and electroweak symmetry breaking scales can be generated by strongly warped regions \cite{Kachru:2003aw}. However, these regions destabilize the K\"{a}hler potential, and stability can be recovered by introducing non-hermitian components to the internal metric. If such non-hermitian parts are treated perturbatively, they can induce NH terms which break SUSY. Previous studies have shown that such mechanisms for SUSY breaking can be controlled by underlying fields \cite{Villadoro:2006ia}; and hence, the NH terms can be induced significantly large but they cannot acquire arbitrary values. A comprehensive consideration of these perturbations \cite{McGuirk:2012sb} has shown that the trilinear NH terms ($T_{u,d,e}^{\prime}$) can be of a similar magnitude as their holomorphic partners ($T_{u,d,e}$). In addition, the supersymmetric fermions (i.e. gauginos and higgsinos) receive supersymmetry breaking masses at similar scales ($M_{i}\sim \mu^{\prime}$) \cite{McGuirk:2012sb}.} 

{These NH terms, especially the trilinear scalar interaction terms can modify the flavour structures of the sfermions; therefore, the sfermions can be rotated differently from their fermionic partners. However, in our study, we assume $T_{u,d,e}^{\prime}$ and $T_{u,d,e}$ to be factorized by the relevant Yukawa matrices as $T_{u,d,e} = A_{u,d,e}y_{u,d,e}$ and $T^{\prime}_{u,d,e} = A^{\prime}_{u,d,e}y_{u,d,e}$. With this alignment in the NH terms, the flavour profiles of the SM fermions can be maintained in their supersymmetric scalar partners. Consequently, the squarks are rotated by the standard CKM matrix, while the sleptons remain diagonal in the flavour basis, which prevents generational mixing among the charged sleptons. In this context, our setup ensures that the NH terms do not introduce new sources for CP-violation, EDM etc. Furthermore, one of our previous studies \cite{Nis:2025fxc} has shown that the NH terms mostly reduce the SUSY contributions to anomalous magnetic moment ($g-2$) such that the results can be consistent with the current status of the experimental muon $g-2$ measurements \cite{Aliberti:2025beg,Muong-2:2025xyk}. We have also shown in this study that the magnitudes of the NH terms do not change much through renormalization group (RG) flow.}

%where $\mu^{\prime}$ is the NH term contributing to the Higgsino mass, while $A_{i}^{\prime}$ ($i=u,d,l$) are similar to the holomorphic trilinear scalar couplings. These terms are classified as NH terms such that the SUSY scalars couple to the ``wrong" Higgs doublet via $A_{i}^{\prime}$. Note that the generation indices are suppressed in Eq.(\ref{eq:SSBLag}), but for the third generation they correspond to $i=t,b,\tau$. 

The third term in Eq.(\ref{eq:SSBLag}) with $T_{b}^{\prime}$ couples the sbottom to $H_{u}$, and it contributes through the first diagram by $T^{\prime} \equiv A_{b}^{\prime}y_{b}$. A similar NH contribution can be realized in the second diagram through the stop and $H_{u}$ vertex, but such contributions are suppressed by $\tan\beta$, and thus they are negligible in comparison with the contributions from $A_{t}$. On the other hand, the main NH contribution in the second diagram appears in the Higgsino mass, which turns out to be $m_{\tilde{H}} = \mu + \mu^{\prime}$. Thus, one can summarize the leading NH contributions to $y_{b}$ as follows:

\begin{equation}
\delta_{y_{b}}^{{\rm NH}} \approx \dfrac{g_{3}^{2}}{12\pi^{2}}\dfrac{A_{b}^{\prime} M_{3}\tan\beta}{m_{\tilde{b}}^{2}}+\dfrac{y_{t}^{2}}{32\pi^{2}}\dfrac{\mu^{\prime} A_{t}\tan\beta}{m_{\tilde{t}}^{2}}~~,
\label{eq:deltabNH}
\end{equation}
resulting in a total threshold correction given by $\delta y_{b} = \delta_{y_{b}}^{H} + \delta_{y_{b}}^{NH}$. With the NH terms, the total contribution from the first diagram can be negative even in the cases with $\mu,M_{3} > 0$, when $A_{b}^{\prime}$ is negative. Thus one can accommodate YU with heavy gluinos and relatively light sbottoms. 

Similarly, a negative and large $\mu^{\prime}$ can provide the desired negative threshold corrections to $\delta_{y_{b}}$ through the second diagram; thus YU can be realized in regions with low $\mu$ ($\sim \mathcal{O}(100)$ GeV). This observation makes  YU more interesting from the naturalness point of view (see, for instance, \cite{Baer:2025uzx}). In fine-tuning ($\FT$) discussions, we will employ the following definitions \cite{Baer:2012mv}:

\begin{equation}
\FT\equiv \dfrac{{\rm Max}(C_{i})}{M_{Z}^{2}/2}
\label{eq:delfine}
\end{equation}
with

\begin{equation}
\setstretch{1.5}
C_{i}=\left\lbrace
\begin{array}{l}
C_{H_{d}}=\mid m^{2}_{H_{d}}/(\tan^{2}\beta -1) \mid~, \\
C_{H_{u}}=\mid m^{2}_{H_{u}}\tan^{2}\beta/(\tan^{2}\beta -1) \mid~, \\
C_{\mu}=\mid -\mu^{2}\mid~.
\end{array}
\right.
\label{eq:CFT}
\end{equation}  

{Among $C_{i}$ given above, $C_{H_{d}}$ is suppressed by $\tan\beta$; thus, $\FT$ is determined mainly by $C_{H_{u}}$ and $C_{\mu}$, and the electroweak symmetry breaking (EWSB) requires $\mid-\mu^{2}\mid \simeq \mid m_{H_{u}}^{2} \mid$. Since the NH terms are not involved in the electroweak symmetry breaking explicitly, the fine-tuning definition remains the same. However, the NH terms can still lead to large cancellations through their effects in RGEs for $m_{H_{u}}^{2}$, which can be seen from }

\begin{equation}
\dfrac{d m_{H_{u}}^{2}}{dt} =\left(\dfrac{d m_{H_{u}}^{2}}{dt}\right)_{{\rm MSSM}} +6\left(\dfrac{1}{5}g_{1}^{2} + g_{2}^{2}\right)\mu^{\prime 2} - 6y_{b}^{2}A_{b}^{\prime 2} - 2y_{\tau}^{2}A_{\tau}^{\prime 2}~,
\label{eq:mHuRGE2}
\end{equation}
{where $t=\log(Q)$, and the RGE is written for the run from the GUT scale to the electroweak scale. As seen from Eq.(\ref{eq:mHuRGE2}), $A_{b}^{\prime}$ and $A_{\tau}^{\prime}$ drive $m_{H_{u}}^{2}$ to the negative values (which is favored by the electroweak symmetry breaking), while $\mu^{\prime}$ rather affects in the opposite direction. Their overall effects on $m_{H_{u}}^{2}$ also changes $|-\mu^{2}|$ indirectly. Note that the similar effects from NH terms can be observed in $m_{H_{d}}^{2}$, but since it is suppressed by $\tan\beta$, the NH contributions to the fine-tuning by $m_{H_{d}}^{2}$ become negligible. Another interesting impact from $\mu^{\prime}$ is observed in the Higgsino masses. In the absence of NH terms, low $\FT$ leads to light masses for Higgsinos, which receive negative and strong impacts from the Planck measurements \cite{Planck:2018nkj} on relic abundance of dark matter (DM) and direct detection of DM experiments \cite{Baer:2024hgq}. One can ameliorate the problem of such a strong impact with the NH terms, since the Higgsinos can be heavier with the contributions from $\mu^{\prime}$ \cite{Un:2023wws}.}

%Among $C_{i}$ given above, $C_{H_{d}}$ is suppressed by $\tan\beta$; so $\FT$ is determined mainly by $C_{H_{u}}$ and $C_{\mu}$, where the electroweak symmetry breaking (EWSB) requires $\mid-\mu^{2}\mid \simeq \mid m_{H_{u}}^{2} \mid$ at the electroweak scale. In the absence of NH terms, a low $\FT$ also implies light masses for Higgsinos, which receive negative and strong impacts from the Planck measurements \cite{Planck:2018nkj} on the relic abundance of dark matter (DM) and DM direct detection experiments \cite{Baer:2024hgq}. One can ameliorate this problem with the NH terms, because the Higgsinos can be heavier with the contributions from $\mu^{\prime}$. Since $\mu^{\prime}$ does not take part in EWSB, it does not alter the fine-tuning measurement; and hence, one can also fit the heavy Higgsino masses with low $\FT$ \cite{Un:2023wws}.

\section{Scanning Procedure, Experimental Constraints and Detector Analyses}
\label{sec:scan}

In this section, we briefly describe the fundamental parameter space of the \FTT~ class supplemented with the NH terms. The fundamental parameters and their ranges can be summarized as follows:

\begin{equation}
\setstretch{1.5}
\begin{array}{lcl}
0 \leq & m_{0} & \leq 20 ~{\rm TeV}, \\
0 \leq & M_{2},M_{3} & \leq 5 ~{\rm TeV}, \\
-3 \leq & A_{0}/m_{0} & \leq 3, \\
35 \leq & \tan\beta & \leq 60, \\
0 \leq & m_{H_{d}}, m_{H_{u}} & \leq 20 ~{\rm TeV},\\
& {\rm sgn}(A^{\prime}_{0}), {\rm sgn}(\mu^{\prime}) & =-1,+1~,
\end{array}
\label{eq:psranges}
\end{equation}
where $m_{0}$ is the universal SSB mass term for the scalar supersymmetric matter fields, while $M_{2}$ and $M_{3}$ represent the SSB mass terms for $SU(2)_{L}$ and $SU(3)_{c}$ gauginos. The SSB mass term for the $U(1)_{Y}$ gaugino can be determined by the Pati-Salam relation as 
\begin{equation}
M_{1} = \dfrac{3}{5} M_{2} + \dfrac{2}{5} M_{3}.
\end{equation}
The parameter $A_{0}$ stands for the universal SSB term for the trilinear scalar interactions, and it is varied with respect to its ratio to $m_{0}$. Its range is restricted as given in Eq.(\ref{eq:psranges}) to avoid CCB vacua \cite{Ellwanger:1999bv,Camargo-Molina:2013sta}. Furthermore, $\tan\beta$ is the ratio of the VEVs of $H_{u}$ and $H_{d}$ defined as $v_{u}/v_{d}$, and $m_{H_{u,d}}$ stand for the SSB mass terms for the Higgs doublets.

In our setup, the NH terms are assumed to be generated in a scale-independent way through the SUSY breaking perturbations on $D-$branes. Such perturbations induce the NH terms such that $|\mu^{\prime}| \sim |M_{i}|$ and $|A_{0}^{\prime}| \sim |A_{0}|$, as discussed in the previous section. {In our scans we set $|\mu^{\prime}| = {\rm max}(M_{2},M_{3})$ and $|A_{0}^{\prime}| = |A_{0}|$ to feed the boundary conditions numerically.} Thus, the NH terms do not enlarge the set of fundamental parameters of the models. However, we allow potential phase differences between the NH terms and their holomorphic partners by randomly assigning them positive and negative signs in our scans.

We perform random scans spanned by the fundamental parameters listed in Eq.(\ref{eq:psranges}) by employing the Metropolis-Hastings algorithm \cite{Baer:2008jn,Belanger:2009ti}. The random input values are passed to SPheno-4.0.4 \cite{Porod:2003um,Goodsell:2014bna} numerical calculation package, which was obtained with SARAH \cite{Staub:2008uz,Staub:2015iza}. SPheno first runs the 1-loop RGEs for the Yukawa and gauge couplings from $M_{Z}$ to $\mgut$, which is determined by the unification condition as $g_{1} = g_{2}\simeq g_{3}$, where $g_{1}$, $g_{2}$ and $g_{3}$ are the gauge coupling for $U(1)_{Y}$, $SU(2)_{L}$ and $SU(3)_{C}$, respectively. Once $\mgut$ is determined all the SSB parameters are imposed and SPheno evolves masses and couplings back to $M_{{\rm SUSY}}$ through the 2-loop RGEs, where $M_{{\rm SUSY}} \equiv \sqrt{m_{\tilde{t}_{L}}m_{\tilde{t}_{R}}}$ is the scale at which the supersymmetric particles decouple from the spectrum. SPheno also controls the perturbativity in these runs by restricting the couplings as $\lambda_{i}\leq \sqrt{4\pi}$. In addition, SPheno excludes the solutions if their RGEs encounter Landau pole. {Even though they can still be compatible with the perturbativity limits, solutions, within which large loop contributions are observed, require a careful treatment. In such solutions there might exist significant cancellations arising at the three-loop level \cite{Baer:2021tta}, but they are neglected in the calculations performed by SPheno. Thus, large loop contributions might reduce the reliability of the solutions.} Furthermore, we accept only the solutions which do not yield negative mass-squared values for the scalar particles and are consistent with radiative EWSB (REWSB), which fixes the bilinear Higgs mixing term $\mu$ with $m_{H_{u}}$, $m_{H_{d}}$ and $\tan\beta$ to be consistent with EWSB. In addition, we require all solutions to yield one of the MSSM neutralinos to be LSP to ensure a suitable DM candidate in spectrum. In this context, one can consider LSP neutralino as accounting for the DM implications, and so we transfer the SPheno solutions to micrOMEGAs \cite{Belanger:2018ccd} to add the DM observables in our analyses.

After generating the data, the solutions are subjected to the mass bounds on the supersymmetric particles \cite{ParticleDataGroup:2014cgo,ATLAS:2021twp,ATLAS:2020syg,ATLAS:2022rcw} and the Higgs boson \cite{ATLAS:2012yve,CMS:2012qbp,CMS:2013btf}, constraints from the combined results for rare $B-$meson decays \cite{CMS:2020rox,Belle-II:2022hys,HFLAV:2022esi}, and the latest Planck Satellite measurements \cite{Planck:2018nkj} on the DM relic abundance successively to constrain the LSP neutralino. The list given below summarizes the experimental constraints employed in our analyses:
\begin{equation}
\setstretch{1.8}
\begin{array}{l}
m_h  = 123-127~{\rm GeV}\\
m_{\tilde{g}} \geq 2.1~{\rm TeV}~(1400~{\rm GeV}~{\rm if~it~is~NLSP})\\
1.95\times 10^{-9} \leq{\rm BR}(B_s \rightarrow \mu^+ \mu^-) \leq 3.43 \times10^{-9} \;(2\sigma) \\
2.99 \times 10^{-4} \leq  {\rm BR}(B \rightarrow X_{s} \gamma)  \leq 3.87 \times 10^{-4} \; (2\sigma) \\
0.114 \leq \Omega_{{\rm CDM}}h^{2} \leq 0.126 \; (5\sigma)~.
\label{eq:constraints}
\end{array}
\end{equation}

In addition to these constraints we define the following parameter to measure YU as

\begin{equation}
R_{tb\tau} = \dfrac{{\rm max}(y_{t},y_{b},y_{\tau})}{{\rm min}(y_{t},y_{b},y_{\tau})}~.
\label{eq:rtbtau}
\end{equation}

{The solutions with exact YU ($y_{t}=y_{b}=y_{\tau}$) correspond to $R_{tb\tau}=1$. However, we allow $10\%$ deviation to take into account the unknown threshold corrections to the Yukawa couplings at $\mgut$ due to the symmetry breaking \cite{Hisano:1992jj,Chkareuli:1998wi}. Thus we identify the YU solutions with the condition $R_{tb\tau} \leq 1.1$. The YU condition is employed in optimization of our scans by maximizing the following function} 

\begin{equation}
P = \exp\left(-\dfrac{R_{tb\tau}-1}{2\sigma_{YU}}\right),
\label{eq:Bayes2}
\end{equation}
{where $\sigma_{YU} = 0.1$, since we allow $10\%$ deviation from the exact YU. However, this optimization strategy does not impose any condition in acceptance of solutions in our results. In other words, the solutions are accepted even if they lead to $R_{tb\tau} \gg 1.1$. We include such solutions in the data to reveal the impact of YU on the parameter space.}

%An exact YU ($y_{t}=y_{b}=y_{\tau}$) solution corresponds to $R_{tb\tau}=1$. However, we allow a $10\%$ deviation to take into account the unknown threshold corrections to the Yukawa couplings at $\mgut$ due to symmetry breaking \cite{Hisano:1992jj,Chkareuli:1998wi}. Thus we identify the YU solutions with the condition $R_{tb\tau} \leq 1.1$.

\begin{figure}[h!]
\centering
\includegraphics[scale=0.35]{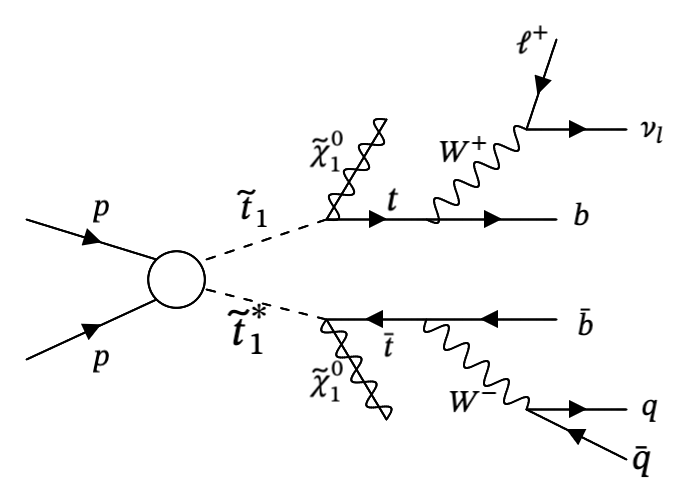}
\caption{The topology for the signal processes considered in our analyses.}
\label{fig:sigtop}
\end{figure}

At the end of our analyses, we also discuss several benchmark scenarios to consider possible signals which can be probed in collider experiments. We select some solutions which are consistent with the constraints listed in Eq.(\ref{eq:constraints}). In such analyses, we consider stop pair production in which each stop subsequently decays into an LSP neutralino associated with a top quark. The process is diagrammatically shown in Figure \ref{fig:sigtop}. Note that we consider the semi-leptonic final states in our analyses due to the detector efficiency in $b-$tagging, which cannot exceed $70\%$ \cite{ATLAS:2022miz}. While we can keep $b-$tagging efficiency at about $70\%$ in semi-leptonic decays, it would be reduced to about $50\%$ in fully hadronic final states.

In the analyses for the collider implications, we employ MadGraph5\_aMC@NLO (v3.6.3) \cite{Alwall:2014hca} to generate the signal and relevant background events at the parton level. Following the analyses in Ref.\cite{CMS:2025ttk}, we set the five-flavor-scheme (5FS) and use NNPDF23LO1 \cite{NNPDF:2017mvq} at this step. The generated events are transferred to Pythia 8.3 \cite{Sjostrand:2014zea} to complete the final states by parton showering, hadronization, and decay of unstable particles. The detector response is also included in our simulations with the use of Delphes-3.5.0 \cite{deFavereau:2013fsa}. Since we expect heavy stops which are more likely beyond the reach of the current experiments, we simulate similar analyses to those reported in Ref.\cite{CMS:2025ttk} for the FCC-hh with a 100 TeV Center of Mass (COM) energy, by employing the  \texttt{FCChh\_II.tcl} card within Delphes, which is generated suitably with the official design of FCC-hh. We discuss our results at this step in terms of the statistical significance for the signal events, which is calculated as follows \cite{Cowan:2010js, Kumar:2015tna}:

\begin{equation}
\mathcal{Z} =
\sqrt{
  2\left[
    (S+B)\,\ln\left(
      1 + \dfrac{S}{B}
    \right)
    - S
  \right]
},
\label{eq:SS_0}
\end{equation}
where $S$ represents the number of events for the signal, while $B$ stands for the total background processes.

\section{Fundamental Parameter Space of YU}
\label{sec:FPUYU}

In this section, we evaluate the impact of the NH terms on the fundamental parameter space of YU and the fine-tuning measure. In the first part, we will provide a comparison between results from two independent sets of scans. One data-set called ``Holomorphic" does not take the NH contributions into account, while the other (denoted by NH) incorporates the contributions from these terms in numerical calculations. Both sets are optimized to have statistically well-distributed YU solutions. After the comparison, we will discuss the differences in terms of the NH threshold corrections defined in Eq.(\ref{eq:deltabNH}). We will also include briefly the Higgsino-like LSP solutions and their DM implications.

\subsection{Comparison between Holomorphic and NH Cases}
\label{subsec:comparison}

\begin{figure}[t!]
\centering
\includegraphics[scale=0.6]{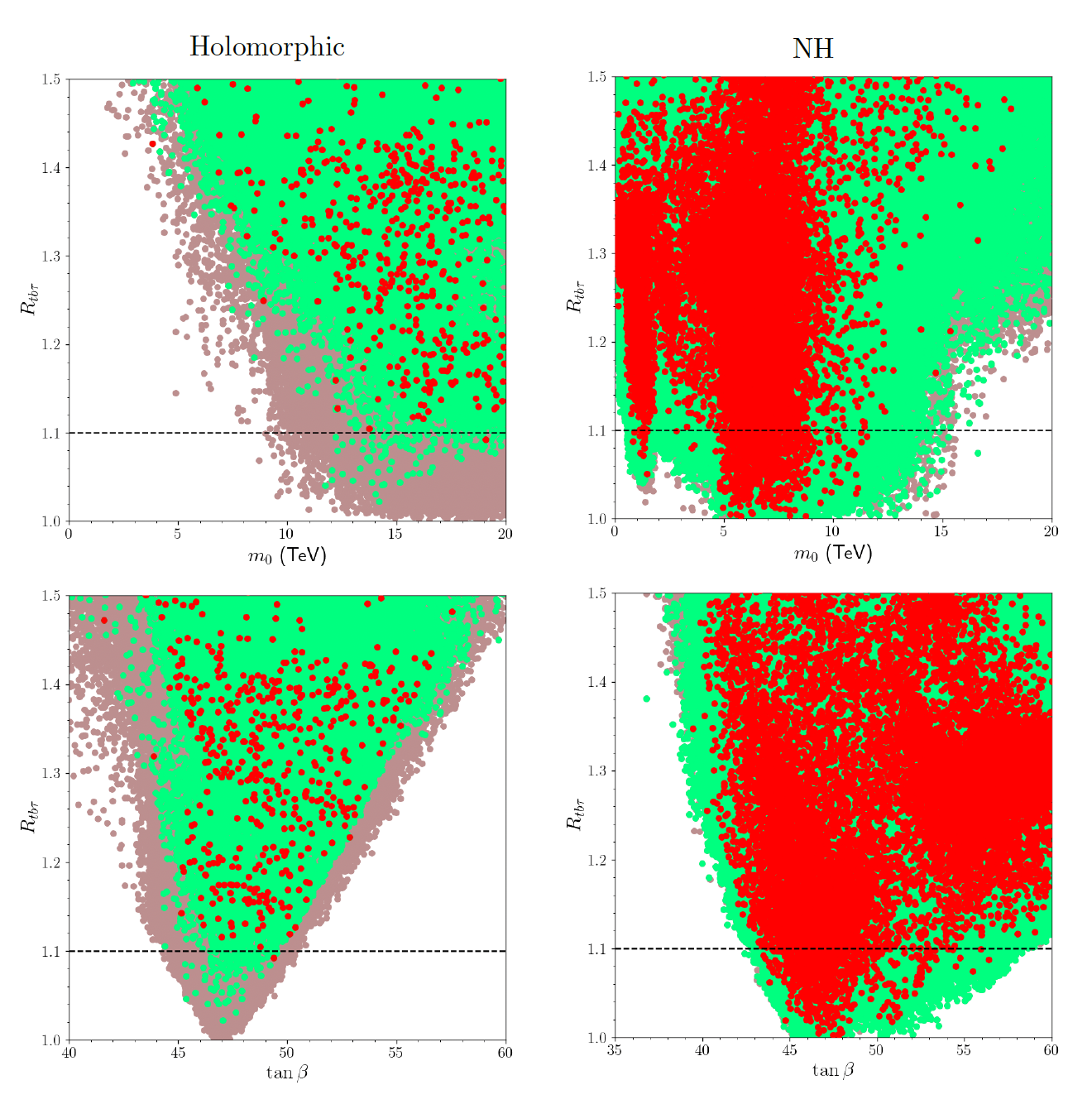}
\caption{The results for YU in correlation with $m_{0}$ (top) and $\tan\beta$ (bottom) for the holomorphic (left) and NH (right) cases. All the solutions are compatible with REWSB and LSP neutralino condition. The green points satisfy the mass bounds and constraints from rare $B-$meson decays. The red points form a subset of green, and they represent the relic density solutions compatible with the Planck measurements within $5\sigma$. The horizontal dashed lines indicate the solutions with $R_{tb\tau} = 1.1$, and the points below these lines are identified as the YU solutions.}
\label{fig:YUm0tanb}
\end{figure}

\begin{figure}[t!]
\centering
\includegraphics[scale=0.6]{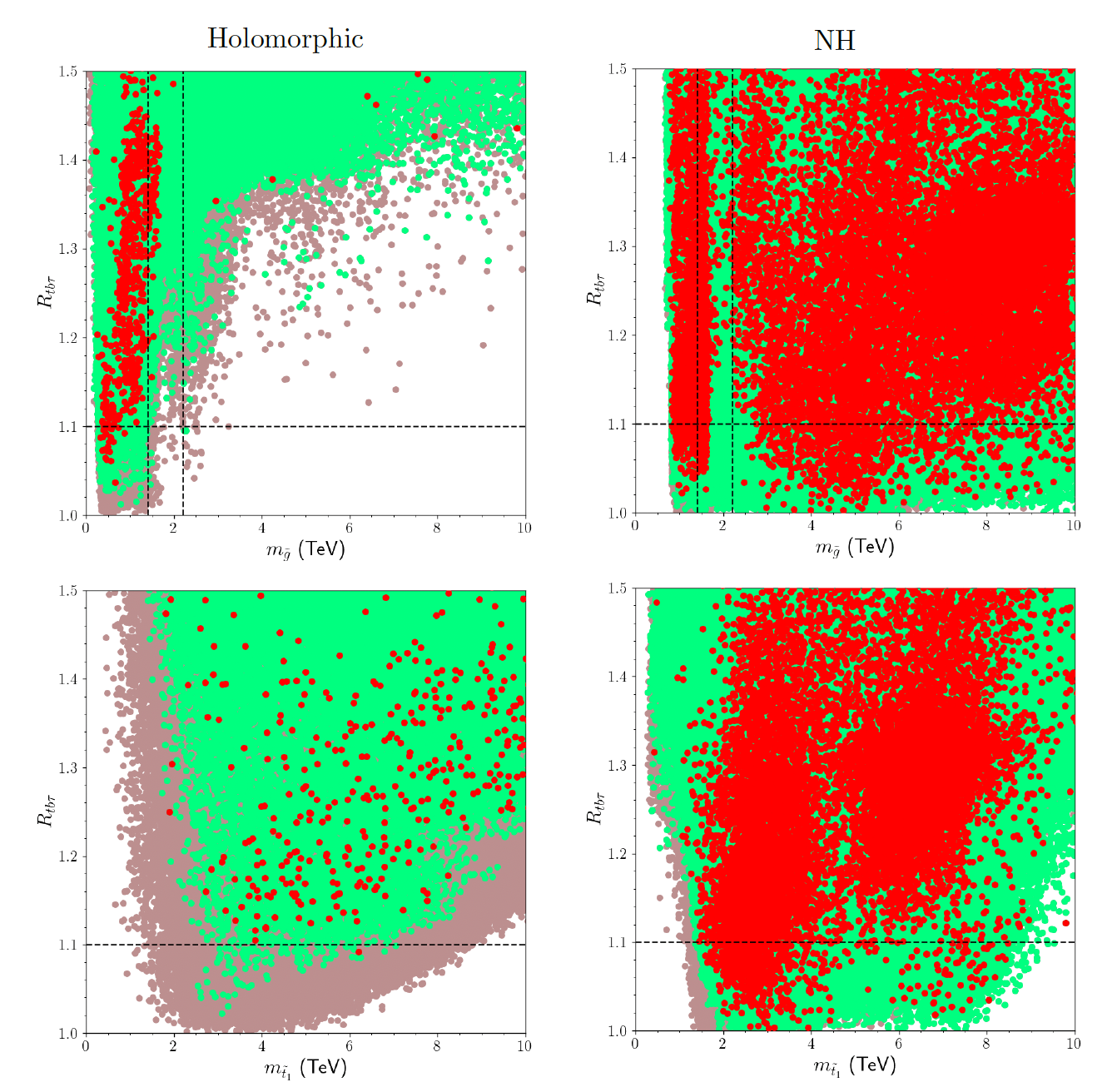}
\caption{The plots in the $R_{tb\tau}-m_{\tilde{g}}$ (top) and  $R_{tb\tau}-m_{\tilde{t}_{1}}$ (bottom) planes for the holomorphic (left) and NH (right) cases. The color coding and the dashed line represent the solutions as described in Figure \ref{fig:YUm0tanb}. The gluino mass bounds are not applied to the top planes, while these bounds are represented with vertical dashed lines.}
\label{fig:YUSt1Glu}
\end{figure}

\begin{figure}[t!]
\centering
\includegraphics[scale=0.6]{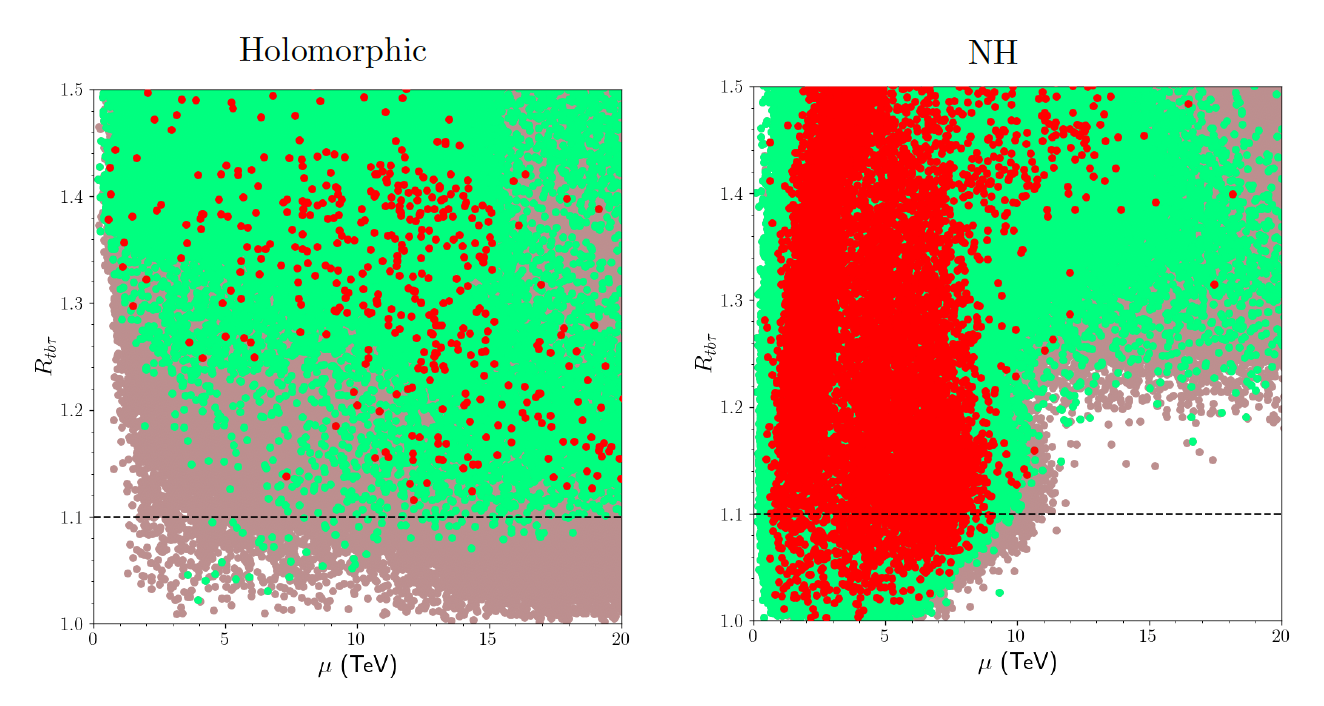}
\caption{The plots in the $R_{tb\tau}-\mu$ in the holomorphic (left) and NH (right) cases. The colors and the dashed lines represent the solutions as described in Figure \ref{fig:YUm0tanb}}
\label{fig:YUmu}
\end{figure}

We display the YU solutions parameterized by $R_{tb\tau}$ in correlation with the universal scalar mass, $m_{0}$ (top) and $\tan\beta$ (bottom) in Figure \ref{fig:YUm0tanb}. The left panels show fundamental parameter space of YU in the holomorphic case, while the right panels display the results including the NH contributions. These two datasets are generated independently to accommodate the best possible YU solutions in both cases. All the solutions are compatible with REWSB and the LSP neutralino condition. The green points satisfy the mass bounds and constraints from rare $B-$meson decays. The red points form a subset of green, and they represent the relic density solutions compatible with the Planck measurements within $5\sigma$. The horizontal dashed lines indicate the solutions with $R_{tb\tau} = 1.1$, and the points below these lines are identified as the YU solutions. As mentioned before, the YU condition exerts a strong impact in the low scale implications which can be seen clearly in the $R_{tb\tau}-m_{0}$ plane in the top-left. In the holomorphic case, the YU solutions can be realized only at heavy scales of $m_{0}$ ($\gtrsim 10$ TeV), and demanding the correct relic density of LSP neutralino (red) shifts it to much heavier mass scales. Consequently, these results lead to heavy mass spectra for the scalars at the low scale. On the other hand, when the NH contributions are included one can accommodate the YU solutions in regions where $m_{0}$ is as low as about 1 TeV, as seen from the top-right plane. The strong impact of YU can also be observed in the correlation between $R_{tb\tau}$ and $\tan\beta$ shown in the bottom planes. The YU solutions require a very narrow range ($45 \lesssim \tan\beta \lesssim 50 $), and if one requires a perfect YU solution ($R_{tb\tau} = 1$), it can be realized specifically when $\tan\beta \simeq 47$. This specific value for $\tan\beta$ is determined mostly by the mass ratio of the  top and bottom quarks. The NH contributions loose this strong correlation between YU and $\tan\beta$ as seen from the bottom-right plane. YU can be accommodated in a wider range of $\tan\beta$ as $43 \lesssim \tan\beta \lesssim 60$. Moreover, the exact YU solutions can also be realized abundantly in the range $45 \lesssim \tan\beta \lesssim 52$.

The impact of the results discussed in the top planes of Figure \ref{fig:YUm0tanb} can immediately be seen in the mass spectrum compatible with the YU condition. Figure \ref{fig:YUSt1Glu} displays our results in the $R_{tb\tau}-m_{\tilde{g}}$ (top) and  $R_{tb\tau}-m_{\tilde{t}_{1}}$ (bottom) planes for the holomorphic (left) and NH (right) cases. The color coding and the horizontal dashed lines represent the solutions as described in Figure \ref{fig:YUm0tanb}. The gluino mass bounds are not applied to the top planes, while these bounds are represented with vertical dashed lines. As discussed in Section \ref{sec:NHYU}, the gluino should be light to keep the positive threshold corrections small in the holomorphic case, and the $R_{tb\tau}-m_{\tilde{g}}$ plane in the top-left panel shows its consequence directly. The solutions in the regions between the two vertical dashed lines are allowed only when the gluino is NLSP. The results show that consistent gluino masses can be barely fit with the YU condition (green), and these solutions cannot be compatible with the desired relic density for the LSP neutralino to be accounted for in DM observations. YU becomes significantly worse in the region of heavy gluino ($R_{tb\tau} \gtrsim 1.2$). The positive threshold corrections from the first term in Eq.(\ref{eq:deltab}) are so large in this region that the Yukawa couplings cannot unify. On the other hand, these large positive contributions can be compensated by the negative NH contributions such that YU can be accommodated with heavy gluino solutions as shown in the top-right plane of Figure \ref{fig:YUSt1Glu}. Similarly, in the holomorphic case, the stop can only be as light as about 2.2 TeV, while with the NH contributions its mass can be lowered to about 1.5 TeV as shown in the bottom planes. Even though the lowest mass scales for stops are close in both cases, the main impact can be seen in the DM implications. In the holomorphic case, the correct relic density of LSP neutralino (red) can be obtained when the stop weighs about 6 TeV, while the NH case abundantly fits the light stop solutions with the suitable DM relic density. 

We conclude the comparison between the holomorphic and NH cases with the allowed ranges for $\mu$ in Figure \ref{fig:YUmu} with plots in the $R_{tb\tau}-\mu$ planes for the holomorphic (left) and NH (right) cases. The color coding and the dashed lines represent the solutions as described in Figure \ref{fig:YUm0tanb}. As discussed before, YU favors large $\mu$ values in the holomorphic case, and the left panel shows that one can accommodate the YU solutions in regions with $\mu \gtrsim 4$ TeV. On the other hand, the desired negative contributions to $y_{b}$ can be provided with $\mu^{\prime}$ and $A_{b}^{\prime}$ as given in Eq.(\ref{eq:deltabNH}). With these NH contributions, the $\mu-$term can be even as low as about 100 GeV as shown in the right plane of Figure \ref{fig:YUmu}. Indeed, the NH contributions put an upper bound on $\mu$ such that it cannot be larger than about 10 TeV. Beyond these values, $y_{b}$ starts receiving excessively negative contributions which also break YU.

\subsection{YU, Low Fine-Tuning and Higgsino-like LSP}
\label{subsec:FTHiggsino}

\begin{figure}[t!]
\centering
\includegraphics[scale=0.6]{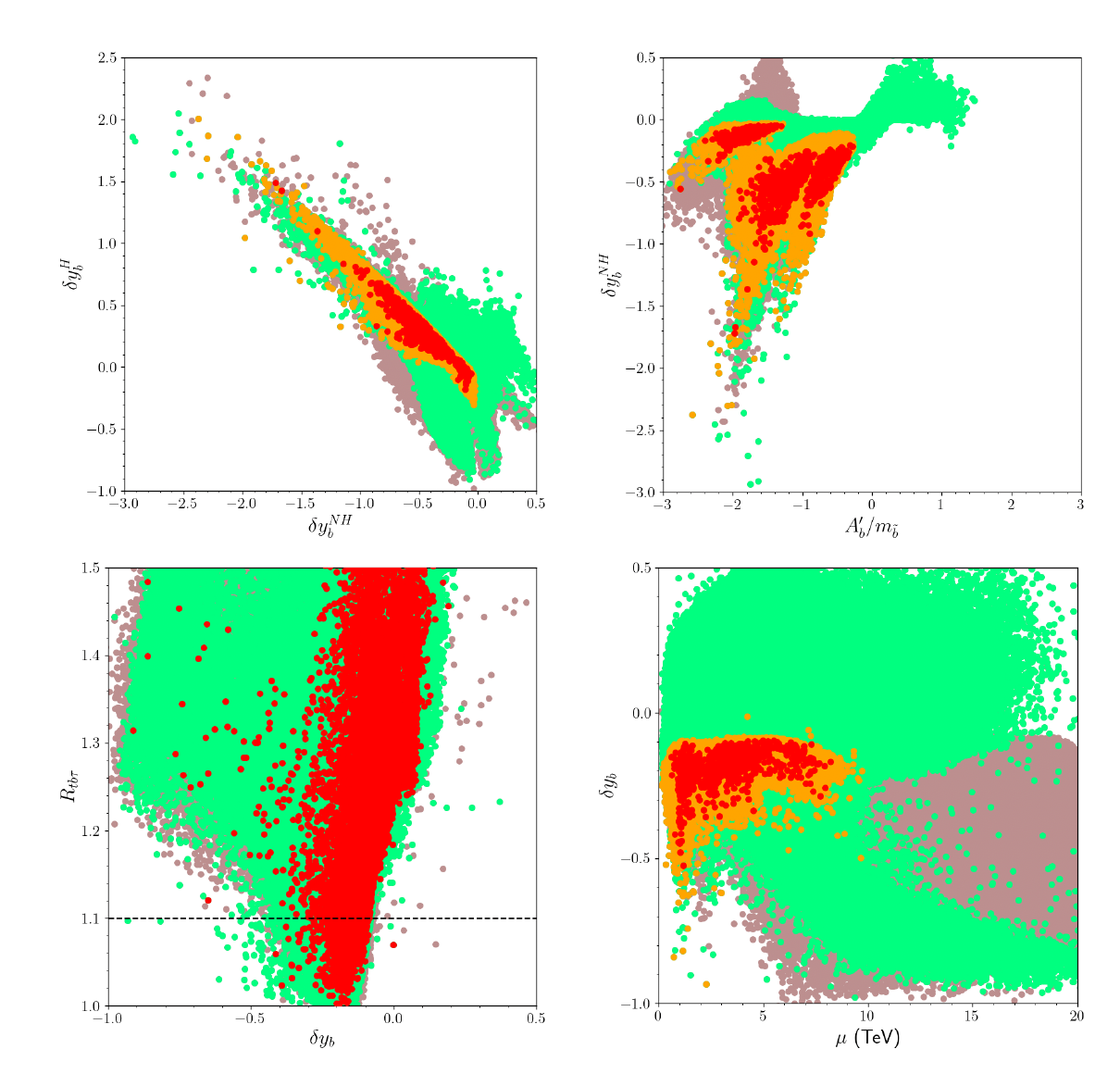}
\caption{The threshold corrections to $y_{b}$ in the $\delta_{y_{b}}^{H}-\delta_{y_{b}}^{NH}$, $\delta_{y_{b}}^{NH}-A_{b}^{\prime}/m_{\tilde{b}}$, $R_{tb\tau}-\delta_{y_{b}}$ and $\delta_{y_{b}}-\mu$ planes. The color coding in the bottom-left plane is the same as in Figure \ref{fig:YUm0tanb}. In the other planes, the green points show the solutions consistent with the mass bounds and constraints from rare $B-$meson decays. The orange points form a subset of green which are compatible with the YU condition. Red solutions on top of orange represent the YU solutions with the relic density allowed by the  Planck measurements within $5\sigma$.}
\label{fig:deltayb}
\end{figure}

\begin{figure}[t!]
\centering
\includegraphics[scale=0.6]{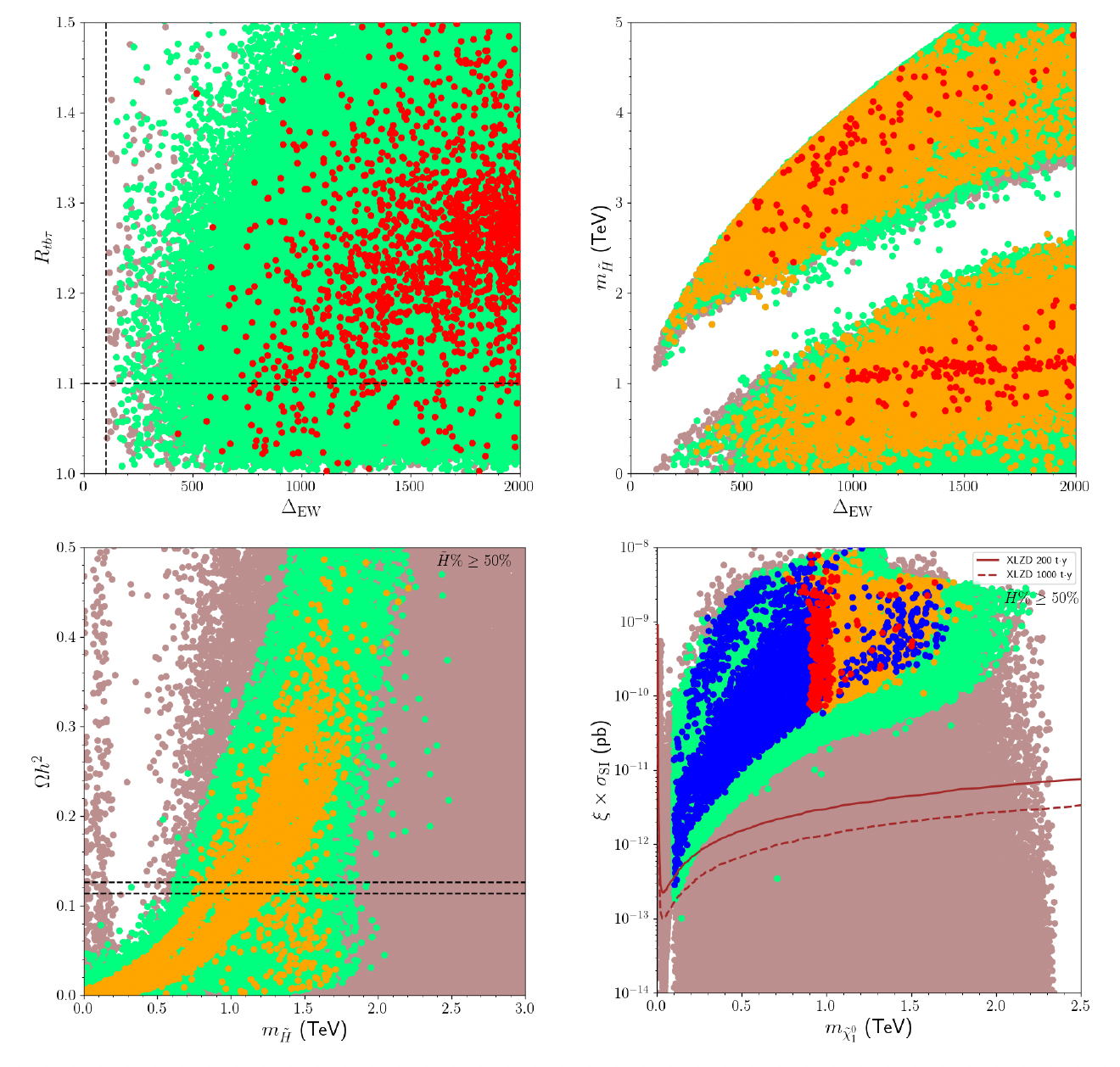}
\caption{The impact of the NH terms in the fine-tuning and DM implications. The color coding in the top-left plane is the same as in Figure \ref{fig:YUm0tanb}, while the other planes follow the color coding as described for the top planes in Figure \ref{fig:deltayb}. {The bottom panels display only the solutions in which the Higgsinos form DM by $50\%$ and more. The horizontal dashed lines in the $\Omega h^{2}-m_{\tilde{H}}$ plane represent the current measurement of Planck satellite within $5\sigma$. The solid (dashed) curves in the bottom-right panel represents the current (future projected) sensitivity of the XENON experiment \cite{XENON:2020kmp}. In addition, the solutions with low relic density of the LSP neutralino are displayed with blue points in this panel.}}
\label{fig:FTDM}
\end{figure}

As is seen above, the NH contributions can significantly expand the parameter space of YU. This is a direct impact of the NH terms through their threshold corrections to $y_{b}$ given in Eq.(\ref{eq:deltabNH}). Indeed, they can completely dominate the threshold corrections such that the YU solutions can be realized almost independently of the holomorphic contributions. We display these corrections in the $\delta_{y_{b}}^{H}-\delta_{y_{b}}^{NH}$, $\delta_{y_{b}}^{NH}-A_{b}^{\prime}/m_{\tilde{b}}$, $R_{tb\tau}-\delta_{y_{b}}$ and $\delta_{y_{b}}-\mu$ planes in Figure \ref{fig:deltayb}. The color coding in the bottom-left plane is the same as in Figure \ref{fig:YUm0tanb}. In the other planes, the green points show the solutions consistent with the mass bounds and constraints from rare $B-$meson decays. The orange points form a subset of green which are compatible with the YU condition. Red solutions on top of orange represent the YU solutions with the relic density allowed by the  Planck measurements within $5\sigma$. The top-left plane compares the holomorphic and NH contributions based on Eqs.(\ref{eq:deltab} and \ref{eq:deltabNH}). As is seen from the results in this plane, the holomorphic contributions can be positive in most of the YU solutions (orange) and even can be as large as about 2. Such large positive contributions are compensated by large negative NH contributions ($\sim -2.5$). The need for such large negative NH contributions require negative $A_{b}^{\prime}$ as seen in the $\delta_{y_{b}}^{NH}-A_{b}^{\prime}/m_{\tilde{b}}$ plane. The magnitude of $A_{b}^{\prime}$ is about three times as much as the sbottom mass. 

The bottom planes of Figure \ref{fig:deltayb} display the total threshold corrections to $y_{b}$ ($\delta y_{b}$) to accommodate the YU solutions. The $R_{tb\tau}-\delta_{y_{b}}$ plane shows that the total corrections to $y_{b}$ mostly happen to be negative in the range $-1 \lesssim \delta y_{b} \lesssim 0$ (green points under the horizontal dashed line). The relic density constraint (red) shrinks this region to $-0.5 \lesssim \delta y_{b} \lesssim 0$. {It should be noted that the threshold corrections to $y_{b}$ can reach the values as large as about $-1$, but as discussed earlier, such large corrections should be reconsidered with higher-order loop corrections, which might lead to significant cancellations.} Finally we also display the possible ranges for the $\mu-$term which makes the discussions of fine-tuning and the Higgsino-like LSP relevant to our results. The YU solutions (orange) in the $\delta_{y_{b}}-\mu$ plane can be compatible with low values of $\mu$ ($\gtrsim 100$ GeV), while they can also be as large as about 10 TeV. 

We convert the values of $\mu$ to the fine-tuning measure by following the definitions in Eqs.(\ref{eq:delfine} and \ref{eq:CFT}) in Figure \ref{fig:FTDM}. The impact of the NH terms in the fine-tuning and DM implications. The color coding in the top-left plane is the same as in Figure \ref{fig:YUm0tanb}, while the other planes follow the color coding as described for the top planes in Figure \ref{fig:deltayb}. {The bottom panels display only the solutions in which the Higgsinos form DM by $50\%$ and more. The horizontal dashed lines in the $\Omega h^{2}-m_{\tilde{H}}$ plane represent the current measurement of Planck satellite within $5\sigma$. The solid (dashed) curves in the bottom-right panel represents the current (future projected) sensitivity of the XENON experiment \cite{XENON:2020kmp}. In addition, the solutions with low relic density of the LSP neutralino are displayed with blue points in this panel.} As is seen from the top-left plane, the NH contributions can significantly improve the fine-tuning results. The YU solutions can be realized with $\FT$ as low as about 100, while the relic density constraint (red) can lift the lower bound on $\FT$ to about 400. As discussed before, the Higgsino mass does not have to be light for such solutions because of the contributions from $\mu^{\prime}$, and the $m_{\tilde{H}}-\FT$ plane shows that the Higgsinos can be as heavy as about 1 TeV in this region. However, despite the significant improvement in $\FT$, the Higgsinos receive a strong impact from the Planck measurements and DM direct detection experiments, when they form LSP. The $\Omega h^{2}-m_{\tilde{H}}$ plane shows that the correct relic density can be satisfied for the Higgsino masses in the range $0.7 \lesssim m_{\tilde{H}}\lesssim 1.5$ TeV, when they participate in the LSP composition more than $50\%$. The lighter Higgsino masses lead to low relic density, and these solutions can still take part in DM observations, but they should be accounted for the testable implications partially. One can reflect their partial contributions in DM observations by rescaling their scattering cross-section as \cite{Belanger:2015vwa}:

\begin{equation}
\setstretch{2.5}
\xi = \left\lbrace \begin{array}{ll}
1 & {\rm for~} 0.114 \leq \Omega h^{2} \leq 0.126~, \\
\dfrac{\Omega h^{2}}{0.12} & {\rm for}~ \Omega h^{2} < 0.114~.
\end{array}\right.
\label{eq:xi}
\end{equation} 

Even with such a rescaling, the Higgsino LSP can still be excluded by the current analyses of the direct detection experiments as shown in the bottom-right plane of Figure \ref{fig:FTDM}. We display only the solutions for which the LSP neutralino is formed by the Higgsinos by more than $50\%$. Despite their rescaled scattering cross-section, the current sensitivity of the XENON experiment (solid curve) is enough to exclude such solutions. The solutions can be compatible with these results only when $m_{\tilde{\chi}_{1}^{0}} \gtrsim 100$ GeV, but these solutions yield quite negligible relic density ($\Omega h^{2} \lesssim 10^{-3}$) such that their contributions to the DM observations are negligible.

\begin{figure}[h!]
\centering
\includegraphics[scale=0.4]{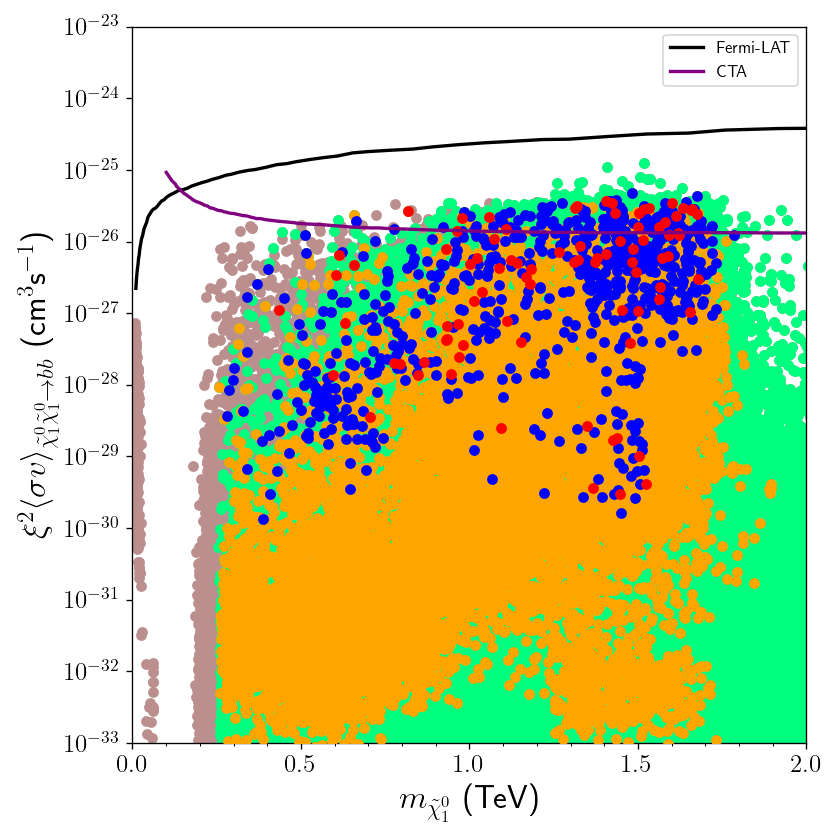}%
\includegraphics[scale=0.4]{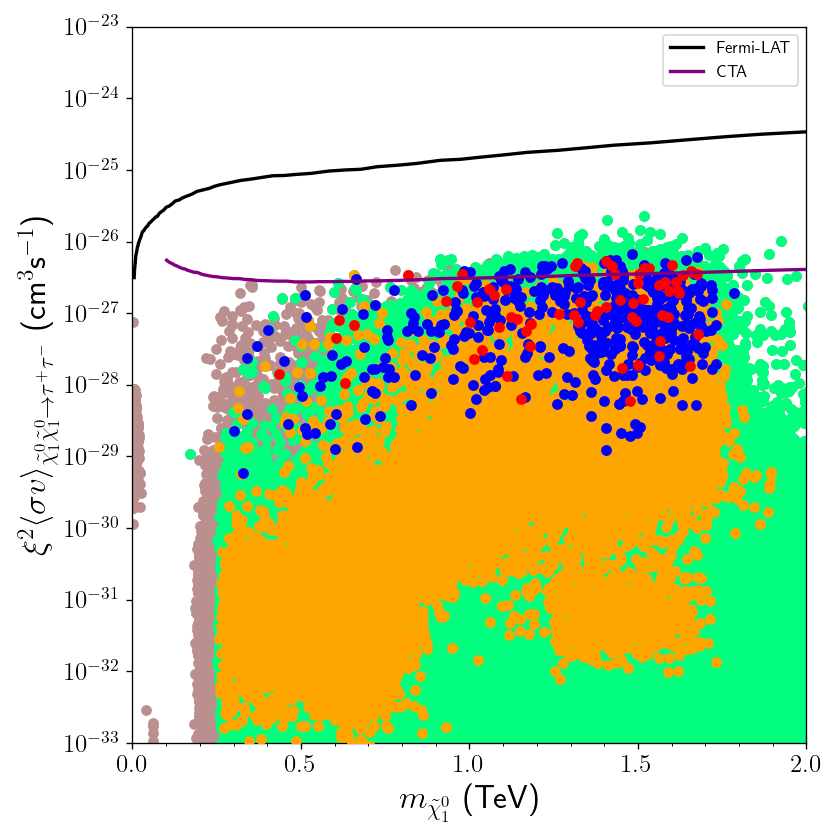}
\caption{Results for the indirect detection of DM in its annihilation channels into a pair of bottom quarks (left) and $\tau-$leptons (right). All points are compatible with the REWSB and LSP neutralino conditions. Green points satisfy the mass bounds and constraints from rare $B-$meson decays. Orange points form a subset of green, and they display the YU solutions. Blue points are the solutions with lower DM relic density, while those compatible with the current Planck measurements are shown in red. The curves represent the current exclusion curves from Fermi-LAT \cite{Fermi-LAT:2017opo}, and Cherenkov Telescope Array (CTA) \cite{CTA:2020qlo} as indicated in the legend. The cross-section values are rescaled with $\xi^{2}$, which is defined in Eq.(\ref{eq:xi})}
\label{fig:inD}
\end{figure}

{One can also consider the implications for indirect detection of DM by considering the annihilation channels into the SM particles. In our analyses, we realize that the DM annihilations into a pair of bottom quarks and $\tau-$leptons can yield relatively large average cross-section which can be tested by the current measurements of the space-based and ground-based gamma-ray observatories. We display our results for these annihilation channels of DM in Figure \ref{fig:inD}. All points are compatible with the REWSB and LSP neutralino conditions. Green points satisfy the mass bounds and constraints from rare $B-$meson decays. Orange points form a subset of green, and they display the YU solutions. Blue points are the solutions with lower DM relic density, while those compatible with the current Planck measurements are shown in red. The curves represent the current exclusion curves from Fermi-LAT \cite{Fermi-LAT:2017opo}, and Cherenkov Telescope Array (CTA) \cite{CTA:2020qlo} as indicated in the legend. The annihilation rates into a pair of bottom quarks can reach to the current sensitivity of the CTA measurements ($\gtrsim 10^{-26}$ cm$^{3}$s$^{-1}$) when the LSP neutralino weighs heavier than about 500 GeV. Even though the current measurements from Fermi-LAT need further upgrades, one can expect the solutions in this region can be re-tested soon in this experiment. Our results reveal that the annihilation rates for $\tau\tau$ channel are slightly lower, but the CTA measurements will exhibit enough sensitivity to probe such solutions. in the same mass scales of LSP neutralino. The annihilation channels discussed in Figure \ref{fig:inD} happen through the Higgs portal in which two LSP neutralinos annihilate into the heavy CP-even ($H$) and/or CP-odd ($A$) Higgs bosons. These results can be interesting for the cases in which the LSP is formed mostly by Higgsinos. However, as discussed in Figure 7, the Higgsino-like LSP solutions receive strong negative impact from the direct detection experiments. Thus, the consistent solutions which can be probed in indirect detection experiments mostly yield Bino-like LSP neutralino solutions. However, our conclusion for the Higgsino DM holds only in the standard scenarios of DM which assume that there is no entropy injection, non-thermal productions for the DM \cite{Arbey:2008kv,Arbey:2009gt,Arcadi:2024jzv} etc. Even in standard scenarios, if the spectrum involves other DM candidates such as axions/saxions, singlinos etc. (see, for instance, \cite{Baer:2023bbn,Hicyilmaz:2021khy}), the mixing of Higgsinos with such states can lead to drastically different implications.}

%In conclusion, the DM observations cannot be accommodated when the Higgsinos form the LSP neutralino by $50\%$ or more. A similar discussion also holds for the Wino-like LSP neutralino solutions. It usually leads to low relic density when it is lighter than about 1 TeV \cite{ATLAS:2024qmx}, and the current experiments can exclude it up to about 2.8 TeV \cite{Fan:2013faa,MAGIC:2016xys,HESS:2016mib}. Thus, the DM observations in our model can be consistently realized, when the LSP neutralino is formed mostly by Bino.

\section{Probing YU through Compressed Spectra: Stop on Neutralino}
\label{sec:stopcoll}

The discussion in the previous section can be concluded by stating that the NH contributions drastically change the fundamental parameter space such that one can consider reanalyzing its low-scale implications. In this context, we continue our discussion with the mass scales of supersymmetric particles and their possible collider probes. In this section, we first discuss the sparticles and their status with respect to the bounds on their mass scales from the current analyses. Since the LHC experiments collide protons, the experimental analyses are expected to be very sensitive to the squarks and gluino. Among them, we discuss the stop separately, since it can be as light as about 1.5 TeV consistent with all the constraints in our results. At the end of this section, we also include detailed collider analyses which can potentially probe stops.

\subsection{SUSY Mass Spectrum}
\label{subsec:SUSYSpec}

\begin{figure}[t!]
\centering
\includegraphics[scale=0.6]{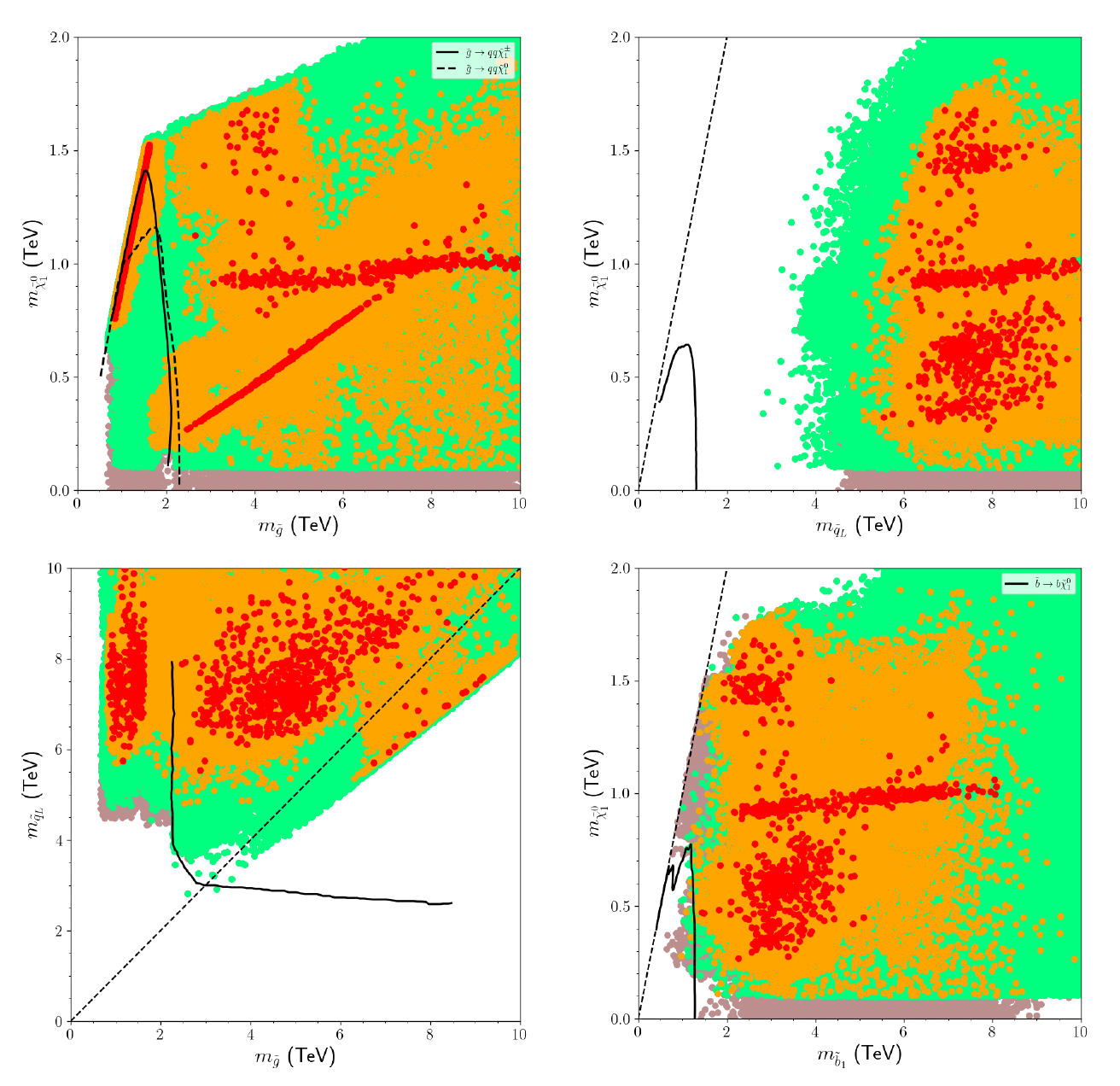}
\caption{The mass spectrum of SUSY particles in the $m_{\tilde{\chi}_{1}^{0}} - m_{\tilde{g}}$, $m_{\tilde{\chi}_{1}^{0}}-m_{\tilde{q}_{L}}$, $m_{\tilde{q}_{L}}-m_{\tilde{g}}$ and $m_{\tilde{\chi}_{1}^{0}}-m_{\tilde{b}_{1}}$ planes. The color coding is the same as in the top planes of Figure \ref{fig:deltayb}. The gluino mass bound is not applied to the results in the left panels. Diagonal lines show the degenerate solutions in masses of the plotted particles. The solid and dashed curves represent the current exclusion bounds as a function of mass from the experimental analyses \cite{ATLAS:2020syg,ATLAS:2021yij}.}
\label{fig:SUSYspec1}
\end{figure}

Figure \ref{fig:SUSYspec1} displays our results for the mass spectra of the supersymmetric particles in the $m_{\tilde{\chi}_{1}^{0}} - m_{\tilde{g}}$, $m_{\tilde{\chi}_{1}^{0}}-m_{\tilde{q}_{L}}$, $m_{\tilde{q}_{L}}-m_{\tilde{g}}$ and $m_{\tilde{\chi}_{1}^{0}}-m_{\tilde{b}_{1}}$ planes. The color coding is the same as in the top planes of Figure \ref{fig:deltayb}. The gluino mass bound is not applied to the results in the left panels. The diagonal lines show the solutions degenerate in the masses of the plotted particles. The solid and dashed curves represent the current exclusion bounds as a function of mass from the experimental analyses \cite{ATLAS:2020syg,ATLAS:2021yij} as indicated in the legends. 

The top-left plane shows our results for the gluino mass in correlation with the LSP neutralino mass. The mass degeneracy between the gluino and LSP neutralino (the orange and red solutions around the diagonal line) can be realized in a gluino mass range of $0.7 \lesssim m_{\tilde{g}}\lesssim 1.5$ TeV, and these solutions can be tested through the gluino pair production which proceeds with the gluino decaying into a pair of quarks along with the LSP neutralino (dashed curve) or a chargino (solid curve) \cite{ATLAS:2020syg}. These analyses can exclude the gluino in this region up to about 1 TeV in its neutralino decays, and even a stronger exclusion can be realized up to about 1.4 TeV through its chargino decays. These results leave a very small window for the NLSP gluino solutions which are more likely tested very soon. In this region, the correct relic density for the LSP neutralino (red points) can be realized through the gluino-neutralino coannihilation scenario; thus the collider analyses will also test indirectly the DM implications as well. The NH contributions, as mentioned before, allow for heavy gluino solutions, and our results show that YU can be realized when the gluino mass lies between about 2.2 TeV and 10 TeV, as well. The current collider experiments are projected to probe these solutions up to about 2.5 TeV, while FCC can significantly upgrade the sensitivity in the gluino searches up to about 6 TeV \cite{Altin:2019veq}. 

The squarks from the first two generations also play a crucial role in direct gluino searches, and the current experimental bounds can exclude them up to about 1.5 TeV through their decays into the LSP neutralino. On the other hand, YU leads to rather heavy spectra for such squarks that they cannot be lighter than about 5 TeV. As is seen from the  $m_{\tilde{\chi}_{1}^{0}}-m_{\tilde{q}_{L}}$ plane, such high mass scales are quite far from the current reach of the experimental analyses (solid curve) \cite{ATLAS:2020syg}. The results from these analyses can be converted into the results shown in the $m_{\tilde{q}_{L}}-m_{\tilde{g}}$ plane. As is seen, almost all the YU solutions (orange and red) accumulate in the region which is consistent with the current experimental results. Note that, the solutions in the light gluino region ($m_{\tilde{g}}\lesssim 1.4$ TeV) are also shown in color in this plane. This is because the gluino mass bound is not applied, which would otherwise exclude almost all the colorful solutions in this region.

The bottom-right plane also shows our results for the sbottom mass scales compatible with YU. After including the NH contributions, one can accommodate sbottoms in SUSY spectra as light as about 1.5 TeV. A small portion of these solutions are already excluded by the current analyses which perform in the region with $m_{\tilde{\chi}_{1}^{0}} \lesssim 800$ GeV \cite{ATLAS:2021yij}. One of the challenges in sbottom analyses is to have final states with $b-$quarks, which can be identified with about 70\% efficiency at the detectors \cite{ATLAS:2022miz}. Thus, when the LSP neutralino weighs heavier than about 800 GeV, the solutions with $m_{\tilde{b}_{1}} \simeq 1.5$ TeV can be consistent with such analyses. As mentioned before, such light sbottom solutions cannot be involved in the YU compatible spectra in the absence of NH contributions. However, our results show that the sbottom searches can also provide a probe for YU, when the NH terms are included.

\begin{figure}[t!]
\centering
\includegraphics[scale=0.4]{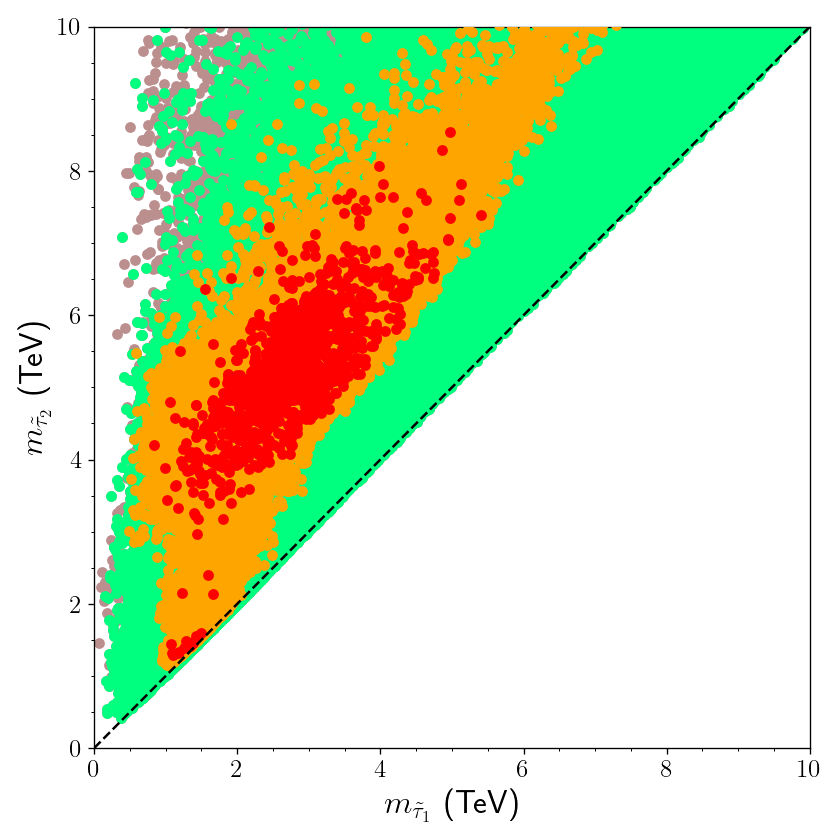}%
\includegraphics[scale=0.4]{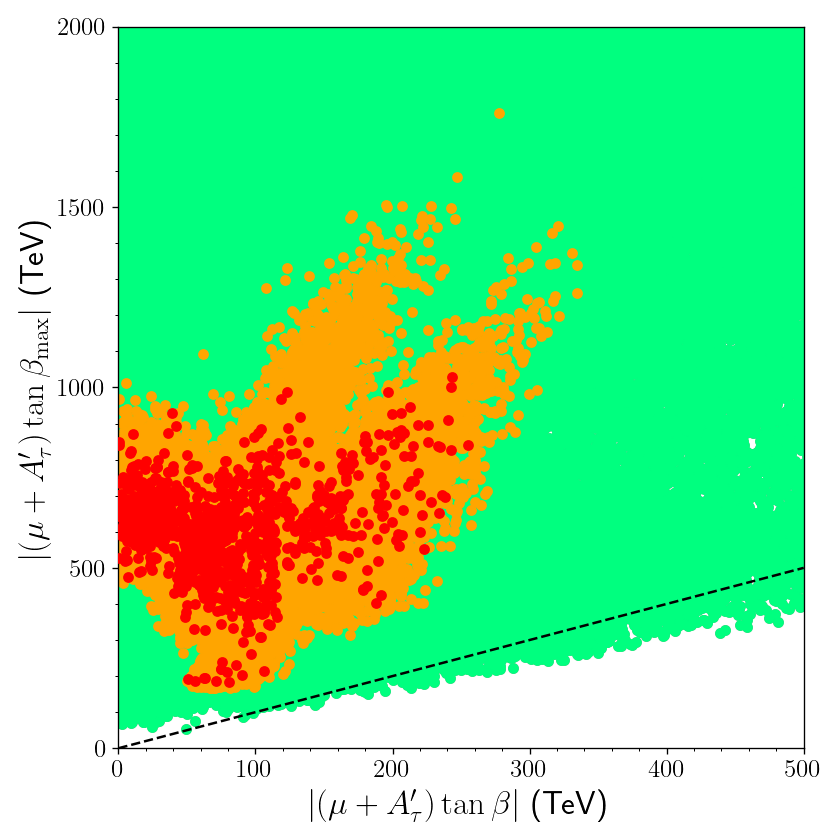}
\caption{Results for the stau masses (left) and the vacuum stability (right) in terms of $|(\mu + A^{\prime}_{\tau})\tan\beta|$ in the parameter region of YU. All the solutions are compatible with the REWSB condition. The green points represent the solutions consistent with the mass bounds and constraints from rare $B-$meson decays. The orange points form a subset of green and they display the YU solutions. The solutions with correct relic density are shown with the red points which are subset of orange. The diagonal line in the left panel indicates the mass degeneracy between $\tilde{\tau}_{1}$ and $\tilde{\tau}_{2}$. In the right panel, the diagonal line shows solutions in which $|(\mu + A^{\prime}_{\tau})\tan\beta|$ takes the possible maximum values based on Eq.(\ref{eq:metasta}).}
\label{fig:metastaus}
\end{figure}

{The SUSY spectra discussed above include the NH contributions through RGEs and the mixing in the sfermion sector. In addition to the detectability of the mass scales in the current collider experiments, these masses and mixing are also important for the vacuum stability \cite{Carena:1995bx,Hisano:2010re,Carena:2011aa,Kitahara:2013lfa,Chigusa:2023mqy}. In MSSM, especially the mixing of the sfermions can destabilize the scalar potential. While the heavy mass scales can keep the squarks from developing non-zero VEVs \cite{Demir:2014jqa,Khalil:2016lgy}, staus can still be crucial for the vacuum stability. The previous analyses have revealed upper bounds on $\mu\tan\beta$ as a function of stau masses. In the presence of the NH terms, $\mu\tan\beta$ should be replaced with $(\mu+A_{\tau}^{\prime})\tan\beta$. Thus, following the analyses in \cite{Kitahara:2013lfa}, the bound on the stau mixing can be expressed as follows:
}
\begin{equation}
\setstretch{2.5}
\begin{array}{rl}
|(\mu + A^{\prime}_{\tau})\tan\beta|  & \lesssim 56.9\sqrt{m_{\tilde{\tau}_{L}}m_{\tilde{\tau}_{R}}} + 57.1(m_{\tilde{\tau}_{L}} + 1.03m_{\tilde{\tau}_{R}})+\dfrac{1.67\times 10^{6}~{\rm GeV}^{2}}{m_{\tilde{\tau}_{L}} + m_{\tilde{\tau}_{R}}} \\ & - (6.41\times 10^{7}~{\rm GeV}^{3})\left(\dfrac{1}{m_{\tilde{\tau}_{L}}^{2}} + \dfrac{0.983}{m_{\tilde{\tau}_{R}}^{2}} \right) - (1.28\times 10^{4}~{\rm GeV})
\end{array}
\label{eq:metasta}
\end{equation}

{Since the NH contributions drive the stau masses to the heavier scales \cite{Nis:2025fxc}, they also lead to higher values for the upper bound on the stau mixing. In this context, one can expect that the vacuum stability is rather improved by the NH terms. We display the results for the stau masses and the vacuum stability in Figure \ref{fig:metastaus}. The color coding is the same as in the top planes of Figure 6. The diagonal line in the left panel indicates the mass degeneracy between $\tilde{\tau}_{1}$ and $\tilde{\tau}_{2}$. In the right panel, the diagonal line shows solutions in which $|(\mu + A^{\prime}_{\tau})\tan\beta|$ takes the possible maximum values based on Eq.(\ref{eq:metasta}). The $m_{\tilde{\tau}_{2}} - m_{\tilde{\tau}_{1}}$ plane shows that $m_{\tilde{\tau}_{1}}$ can be realized as low as about 500 GeV, and they yield large mixing in staus such that $m_{\tilde{\tau}_{2}}$ happens to be larger than about 3 TeV for these solutions. The right plane displays our results for the stau mixing in comparison with its upper bound. As seen from this plane, the stau mixing is realized less than its upper bound for all YU solutions (orange and red).
}

\subsection{Probing Compressed Stop-Neutralino Scenarios}
\label{fig:subsecstop}

\begin{figure}[t!]
\centering
\includegraphics[scale=0.7]{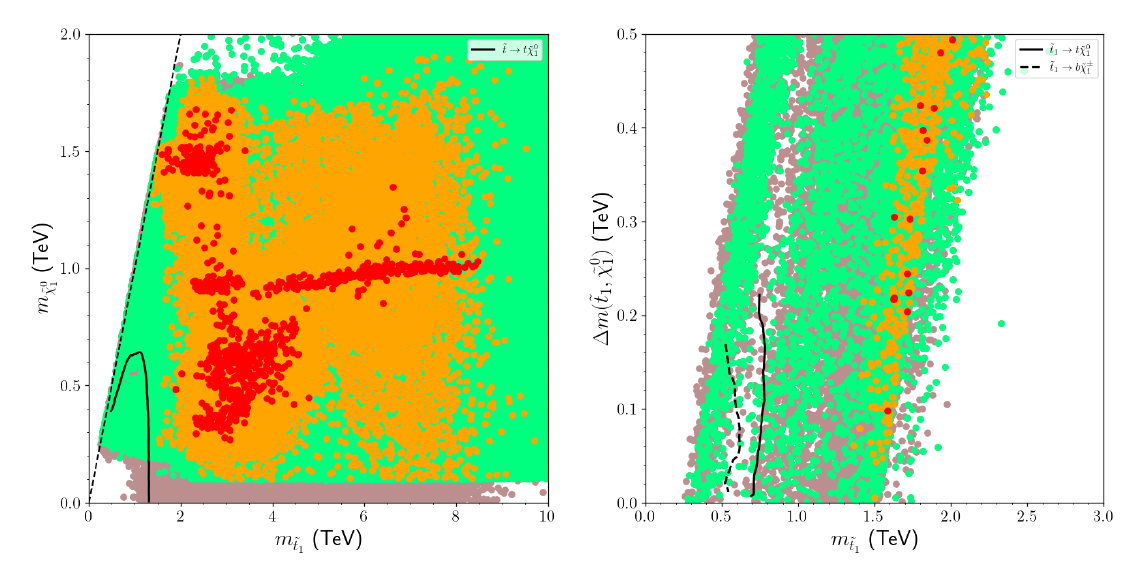}
\caption{The Stop and neutralino masses in the $m_{\tilde{\chi}_{1}^{0}}-m_{\tilde{t}_{1}}$ and $\Delta m(\tilde{t}_{1},\tilde{\chi}_{1}^{0})-m_{\tilde{t}_{1}}$ planes, where $\Delta m(\tilde{t}_{1},\tilde{\chi}_{1}^{0}) \equiv m_{\tilde{t}_{1}}-m_{\tilde{\chi}_{1}^{0}}$. The color coding is the same as in the top planes of Figure \ref{fig:deltayb}. The curves represent the exclusion from the experimental analyses \cite{CMS:2021beq,CMS:2025ttk}.}
\label{fig:SUSYspec2}
\end{figure}

We consider the implications for stops separately from the other sparticles in this section, since they are also of particular interest in detector analyses. Even though these analyses can currently probe the stops up to about 1.2 TeV \cite{CMS:2021beq,CMS:2025ttk}, the main challenge in stop analyses is that the possible signal processes yield final states very similar to those of the main background formed by the top-pair production. These similarities are not limited to the final state particles, but they also include the transverse momenta, missing energy, and angular distributions etc. Thus, suppressing the background processes also kills the signal events significantly \cite{Cici:2016oqr}. In order to realize statistically meaningful number of signal events one needs to consider rather large mass differences between the stop and the LSP neutralino. In this way, the signal events can have considerable statistical significance enhanced with the integrated luminosity. 

Before proceeding to the detailed detector analyses, we present our results for the stop mass scales in Figure \ref{fig:SUSYspec2} in the $m_{\tilde{\chi}_{1}^{0}}-m_{\tilde{t}_{1}}$ and $\Delta m(\tilde{t}_{1},\tilde{\chi}_{1}^{0})-m_{\tilde{t}_{1}}$ planes, where $\Delta m(\tilde{t}_{1},\tilde{\chi}_{1}^{0}) \equiv m_{\tilde{t}_{1}}-m_{\tilde{\chi}_{1}^{0}}$. The color coding is the same as in the top planes of Figure \ref{fig:deltayb}. The curves represent the exclusion bounds from the experimental analyses \cite{CMS:2021beq,CMS:2025ttk}. The YU solutions (orange) around the exclusion curve can yield stops as light as about 1.5 TeV which are slightly beyond the exclusion limit as seen in the $m_{\tilde{\chi}_{1}^{0}}-m_{\tilde{t}_{1}}$ plane. The studies have projected the HL-LHC sensitivity to probe such stop solutions up to about 2.2 TeV \cite{Baer:2023uwo,Baer:2025zqt}, and the FCC-hh experiments will more likely improve the results up to about 3 TeV \cite{Altin:2020qmu}. 

The interesting observation in these results can be realized in the region where the stop and LSP neutralino are nearly degenerate in mass. Such solutions are not present in the holomorphic case \cite{Hussain:2025bxp}. Even though these nearly degenerate stop-neutralino solutions can escape from the sensitivity of the analyses mentioned above, they can be subjected to the analyses over compressed spectra in which the stop does not weigh much heavier than LSP neutralino. Recently released results from these analyses \cite{CMS:2025ttk} are represented with the dashed and solid curves in the $\Delta m(\tilde{t}_{1},\tilde{\chi}_{1}^{0})-m_{\tilde{t}_{1}}$ plane. However, these analyses can probe the stop up to about 700 GeV yet, while the YU solutions appear in the region with $m_{\tilde{t}_{1}}\gtrsim 1.4$ TeV. These solutions are quite beyond the sensitivities of the current collider experiments, but they can be tested in future experiments, which are proposed to operate with much higher center of mass energies such as FCC-hh.

\begin{figure}[h!]
\centering
\includegraphics[scale=0.45]{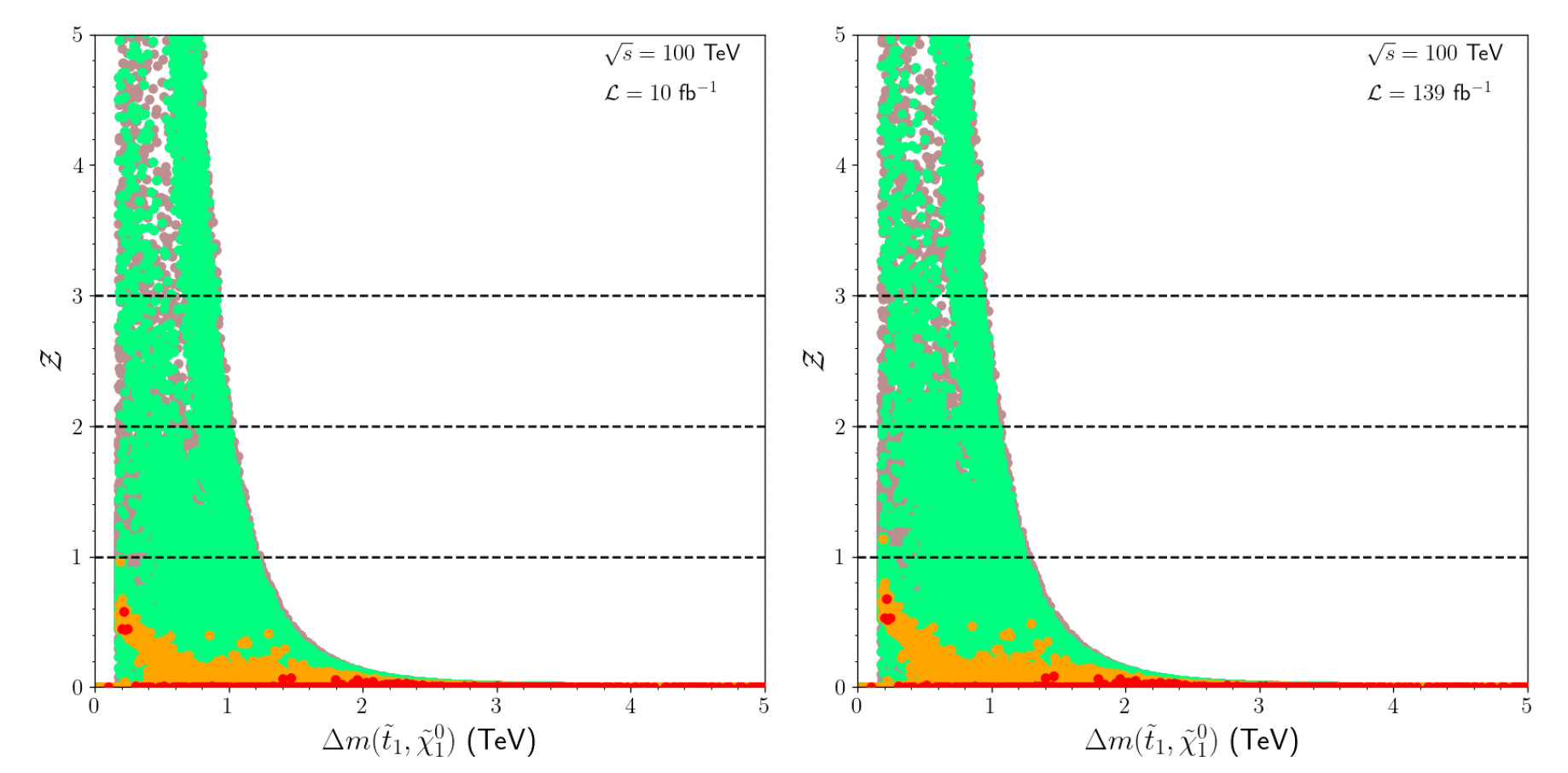}
\caption{The Significance of stop-pair production in correlation with the mass difference between the stop and LSP neutralino at 100 TeV COM with integrated luminosity ($\mathcal{L}$) is 10 fb$^{-1}$ (left) and 139 fb$^{-1}$ (right). In calculation of significance, only the top-quark pair production is considered in the background events such that each stop decays into the LSP neutralino associated with top-quark. The color coding is the same as in the top planes of Figure \ref{fig:deltayb}.  The horizontal dashed lines represent the significance equal to 1, 2 and 3 from bottom to top, respectively.}
\label{fig:SS}
\end{figure}

Figure \ref{fig:SS} displays the significance for the events diagrammatically represented in Figure \ref{fig:sigtop}. The significance calculated as given in Eq.(\ref{eq:SS_0}) projected to the FCC-hh experiments of 100 TeV COM. The color coding is the same as in the top planes of Figure \ref{fig:deltayb}. The horizontal dashed lines represent the significance levels equal to 1, 2 and 3 from bottom to top, respectively. The left panel shows the significance for an integrated luminosity of $\mathcal{L} = 10$ fb$^{-1}$. The YU solutions (orange) can yield stop-pair production events with $\mathcal{Z} \simeq 1$, which can be probed in FCC experiments up to about $68\%$ CL. If one also requires these solutions to be compatible with the Planck measurements on relic abundance of LSP neutralino (red), the maximum significance can be realized at about 0.5. The right panel shows that these results can slightly be improved when the integrated luminosity reaches  139 fb$^{-1}$, yielding $\mathcal{Z} \simeq 1.2$ for the YU solutions (orange) and $\mathcal{Z}\simeq 0.8$ when the YU solutions are consistent with the DM observations (red). 

\begin{table}[h!]
\centering
\setstretch{1.3}\scalebox{0.8}{\begin{tabular}{|c|cccc|}
\hline  & Point 1 & Point 2 & Point 3  & Point 4 \\ \hline
$m_{0}$ &  $ 6218.3 $  &  $ 7210.5 $  &  $ 7434.1 $  &  $ 7729.8 $  \\
$M_{1}$ &  $ 1372.5 $  &  $ 3006.2 $  &  $ 1042.1 $  &  $ 3215.7 $  \\
$M_{2}$ &  $ 747.3 $  &  $ 4625.3 $  &  $ 564.7 $   &  $ 4944.8 $  \\
$M_{3}$ &  $ 2310.2 $  &  $ 577.6 $  &  $ 1758.3 $   &  $ 622.0 $  \\
$m_{H_{d}}$ &  $ 9614.7 $  &  $ 7805.3 $  &  $ 12490.6 $   &  $ 8783.6 $  \\
$m_{H_{u}}$ &  $ 7423.3 $  &  $ 4890.8 $  &  $ 8083.1 $   &  $ 5516.9 $  \\
$A_{0}/m_{0}$ &  $ -1.57 $  &  $ -1.71 $  &  $ -1.25 $   &  $ -1.75 $  \\
$A_{0}^{\prime}/m_{0}$ &  $ -1.57 $  &  $ -1.71 $  &  $ -1.25 $   &  $ -1.75 $  \\
$\tan\beta$ &  $ 46.9 $  &  $ 47.0 $  &  $ 49.1 $  &  $ 43.6 $  \\ \hline
$\mu$ &  $ 3992.4 $  &  $ 7037.5 $  &  $ 3337.1 $   &  $ 7364.3 $  \\
$\mu^{\prime}$ &  $ -1792.4 $  &  $ -3546.1 $  &  $ -1317.8 $  &  $ -3861.0 $  \\
$\Delta_{{\rm EW}}$ &  $ 3.83 \times 10^{3} $  &  $ 1.19 \times 10^{4} $  &  $ 2.68 \times 10^{3} $  &  $ 1.30 \times 10^{4} $  \\ \hline
$m_{h}$ &  $ 126.8 $  &  $ 126.7 $  &  $ 125.9 $  &  $ 126.8 $  \\
$m_{H}$ &  $ 4279.5 $  &  $ 4230.1 $  &  $ 6626.0 $   &  $ 4610.4 $  \\
$m_{A}$ &  $ 4279.5 $  &  $ 4230.1 $  &  $ 6626.0 $   &  $ 4610.4 $  \\
$m_{H^{\pm}}$ &  $ 4283.4 $  &  $ 4234.0 $  &  $ 6630.0 $  &  $ 4613.7 $  \\ \hline
$m_{\tilde{\chi}_{1}^{0}}$,$m_{\tilde{\chi}_{2}^{0}}$ &  $ \cred{619.2} $ ,  $ \cred{649.0} $  &  $ \cred{1409.4} $ ,  $ 3466.6 $  &  $ 473.2 $ ,  $ 502.0 $  &  $ \cred{1509.3} $ ,  $ 3484.8 $  \\
$m_{\tilde{\chi}_{3}^{0}}$,$m_{\tilde{\chi}_{4}^{0}}$ &  $ 2223.7 $ ,  $ 2224.7 $  &  $ 3473.6 $ ,  $ 3966.6 $  &  $ 2039.8 $ ,  $ 2040.9 $ &  $ 3489.5 $ ,  $ 4234.0 $  \\
$m_{\tilde{\chi}_{1}^{\pm}}$,$m_{\tilde{\chi}_{2}^{\pm}}$ &  $ \cred{649.1} $ ,  $ 2225.7 $  &  $ 3466.9 $ ,  $ 3966.6 $  &  $ 502.1 $ ,  $ 2041.9 $  &  $ 3485.0 $ ,  $ 4234.0 $  \\ \hline
$m_{\tilde{g}}$ &  $ 5063.0 $  &  $ \cred{1509.8} $  &  $ 4028.3 $  &  $ 1615.2 $  \\
$m_{\tilde{u}_{1}}$,$m_{\tilde{u}_{2}}$ &  $ 7285.7 $ ,  $ 7401.3 $  &  $ 7178.1 $ ,  $ 7810.9 $  &  $ 7802.9 $ ,  $ 8052.1 $  &  $ 7697.9 $ ,  $ 8366.9 $  \\
$m_{\tilde{t}_{1}}$,$m_{\tilde{t}_{2}}$ &  $ 2972.8 $ ,  $ 3243.4 $  &  $ 2286.7 $ ,  $ 3785.4 $  &  $ 3109.5 $ ,  $ 3730.9 $  &  $ \cred{1713.4} $ ,  $ 3720.6 $  \\ \hline
$m_{\tilde{d}_{1}}$,$m_{\tilde{d}_{2}}$ &  $ 7401.7 $ ,  $ 7426.1 $  &  $ 7302.4 $ ,  $ 7811.4 $  &  $ 8052.5 $ ,  $ 8109.9 $  &  $ 7827.3 $ ,  $ 8367.3 $  \\
$m_{\tilde{b}_{1}}$,$m_{\tilde{b}_{2}}$ &  $ 2944.4 $ ,  $ 3136.9 $  &  $ 2438.6 $ ,  $ 3773.7 $  &  $ \cred{1435.9} $ ,  $ 3128.1 $  &  $ 2218.1 $ ,  $ 3708.7 $  \\ \hline
$m_{\tilde{\nu}_{e}}$,$m_{\tilde{\nu}_{\tau}}$ &  $ 6138.7 $ ,  $ 4353.2 $  &  $ 7703.5 $ ,  $ 5868.0 $  &  $ 7259.6 $ ,  $ 5101.6 $  &  $ 8254.1 $ ,  $ 6499.0 $  \\
$m_{\tilde{e}_{1}}$,$m_{\tilde{e}_{2}}$ &  $ 6139.6 $ ,  $ 6399.8 $  &  $ 7434.4 $ ,  $ 7704.3 $  &  $ 7260.5 $ ,  $ 7756.0 $  &  $ 7965.3 $ ,  $ 8254.9 $  \\
$m_{\tilde{\tau}_{1}}$,$m_{\tilde{\tau}_{2}}$ &  $ 1863.1 $ ,  $ 4354.4 $  &  $ 2307.4 $ ,  $ 5868.9 $  &  $ 2580.2 $ ,  $ 5102.7 $  &  $ 3397.0 $ ,  $ 6499.9 $  \\ \hline
$\Omega h^{2}$ &  $ 0.121 $  &  $ 0.124 $  &  $ 0.121 $  &  $ 0.126 $  \\
$\sigma_{{\rm SI}}$ &  $ 1.57 \times 10^{-12} $  &  $ 1.35 \times 10^{-12} $  &  $ 1.46 \times 10^{-12} $  &  $ 1.73 \times 10^{-12} $  \\
$\sigma_{{\rm SD}}$ &  $ 7.56 \times 10^{-9} $  &  $ 1.14 \times 10^{-9} $  &  $ 1.02 \times 10^{-8} $  &  $ 1.35 \times 10^{-9} $  \\ \hline
$\sigma({\rm Signal})_{14}$ &  $ 1.81 \times 10^{-9} $  &  $ 8.85 \times 10^{-9} $  &  $ 5.68 \times 10^{-12} $  &  $ 7.96 \times 10^{-5} $  \\
$\sigma({\rm Signal})_{100}$ &  $ 1.77 \times 10^{-4} $  &  $ 1.06 \times 10^{-4} $  &  $ 8.60 \times 10^{-7} $  &  $ \cred{1.90 \times 10^{-1}} $  \\
$\mathcal{Z}({\rm Signal})_{100}^{10}$ &  $ 0.0 $  &  $ 0.0 $  &  $ 0.0 $  &  $ 0.5 $  \\ \hline
$R_{tb\tau}$ &  $ \cred{1.01} $  &  $ 1.09 $  &  $ 1.07 $  &  $ 1.09 $  \\ \hline
\end{tabular}
}
\caption{Benchmark scenarios summarizing our findings. In selection the solutions are required to be consistent with all the constraints. All the masses are given in GeV and cross-sections in pb. The subscripts in the cross-section expressions indicate the COM of the collider experiments. The colored values indicate the emphasized features of selected benchmark points.}
\label{tab:benchs}
\end{table}

The significance at this level is preliminary because we assume only that the top-quark pair production forms the SM background, and we do not implement any optimization cut on the kinematic variables. Before proceeding into the detailed analyses and improved significance, we first summarize our findings with four benchmark points in Table \ref{tab:benchs}. We select these points by requiring to be consistent with all constraints applied in our analyses. The masses are given in GeV, while the cross-sections in pb. The subscripts in the cross-sections show the COM of the collider experiments. The statistical significance of the signal events is given for the collisions with 100 TeV COM (shown with the subscript of $\mathcal{Z}$) and 10 fb$^{-1}$ integrated luminosity (shown with superscript of $\mathcal{Z}$). The colored values indicate the emphasized features of the selected benchmark points. Point 1 represents the solutions with almost perfect YU. The mass spectra of such solutions involve typically heavy sparticles, and the Planck measurements on relic density of the LSP neutralino can be realized through chargino-neutralino coannihilation scenario. Point 2 depicts NLSP gluino solutions which are consistent with the current analyses, and it also exemplifies gluino-neutralino coannihilation scenario. Point 3 displays the lightest possible sbottom which is expected to be tested soon in collider analyses. Point 4 represents the stop solutions with maximum possible significance consistent with all the constraints employed in our analyses. The mass difference between the stop and LSP neutralino is about 200 GeV, and it can yield roughly $\mathcal{Z}\simeq 0.5$. 

\subsection{FCC Probe of Compressed Spectrum for Stop and Neutralino}
\label{subsec:stopneut}

The benchmark points given in Table \ref{tab:benchs} can be tested in several analyses as discussed above, but they do not mimic in stop searches except for Point 4. In this section we consider Point 4 as a benchmark scenario, and perform detailed analyses to explore its possible traces in the future collider experiments. In our work we simulate similar analyses to those reported in Ref. \cite{CMS:2025ttk} as briefly described in Section \ref{sec:scan}. 

\begin{table}[h!]
  \centering
  \resizebox{\textwidth}{!}{%
  \begin{tabular}{lcccc}
    \toprule
  Process & $\sigma_{\text{LO}}$ [14 TeV] (pb) & K-Factor [14 TeV] & $\sigma_{\text{LO}}$ [100 TeV] (pb) & K-Factor [100 TeV] \\
    \midrule
  \textbf{Signal} & & & & \\ ($pp \to \tilde{t} \tilde{t}^*$), 
   $ (\tilde{t} \to t \tilde{\chi}_1^0, \bar{\tilde{t}} \to \bar{t} \tilde{\chi}_1^0)$  & $7.96\times 10^{-5}$ & - & $1.90 \times 10^{-1}$& - \\
    \midrule
    \textbf{Backgrounds} & & & & \\
  $pp \rightarrow t \bar{t}$ & $5.056 \times 10^2$ & $1.67$ & $2.469 \times 10^4$ &  $1.5$ \\
   $pp \rightarrow t W^\pm$ & $5.501\times 10^1$& $1.27 $& $2.268\times 10^3$& $1.5$\\
   $pp \rightarrow t \bar{t} V (V = Z, W^\pm)$ & $ 9.36 \times10^{-1}$ & $1.3$ & $4.953 \times 10^1$&  $1.5$ \\
   $pp \rightarrow VV (V = Z, W^\pm)$ &$9.928 \times 10^1$ & $1.35$ & $1.022\times 10^3$ & $1.5$\\
    \bottomrule
  \end{tabular}
  }
  \caption{The signal and relevant background processes at the leading order (LO) and K-factors.}
  \label{tab:bckgs}
\end{table}

\begin{figure}[t!]
%\centering
\includegraphics[scale=0.5]{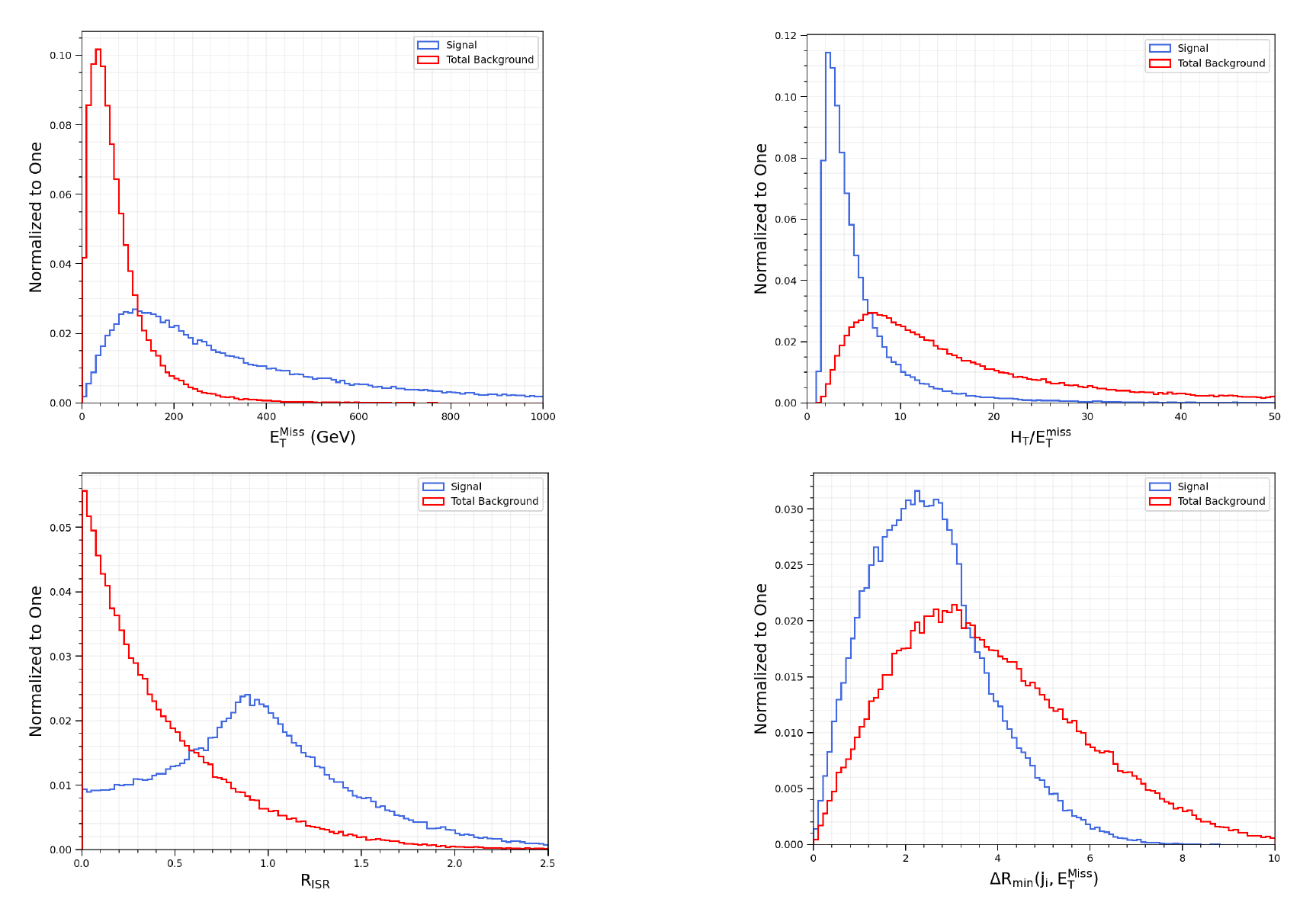}
\caption{The kinematic variables which can distinguish the signal events from those forming the background. The event numbers for the signal and background processes are normalized to 1. $E_{T}^{{\rm Miss}}$ and $H_{T}$ represent the missing and hadronic transverse energies, respectively. $R_{{\rm ISR}}$ denotes the projection of missing transverse energy into the direction of jets emerging from initial state radiation (ISR). Finally $\Delta R_{{\rm min}}$ is the minimum of the angular distance between the variables denoted in parentheses. The distribution for the total background is represented with the red curve, while the blue curve shows for the signal processes.} 
\label{fig:kins}
\end{figure}

Table \ref{tab:bckgs} lists the cross-sections for the signal and relevant background processes at 14 TeV and 100 TeV COM. These values are calculated at the leading order (LO), but the next to LO (NLO) contributions are also included by multiplying them with the K-factors \cite{Kidonakis:2023juy,Frixione:2008yi,Garzelli:2012bn,Grazzini:2019jkl}. Note that an average value is employed for the K-factor when 100 TeV COM is considered \cite{Mangano:2016jyj}. These K-factors are applied only to the background processes. Before considering the relevant kinematic variables, we first apply the following conditions on the events:

\begin{equation}
n_{l} = 1,\hspace{0.3cm} n_{j} \geq 4, \hspace{0.3cm} n_{b} \geq 2,
\label{eq:presel}
\end{equation}
where $n_{l,j,b}$ denote the numbers of leptons, jets and $b-$quarks in the final states, respectively. These conditions can provide suitable pre-selection rules when the semi-leptonic final states are considered as shown in Figure \ref{fig:sigtop}. The further cuts on the kinematic variables are applied on top of these pre-selection rules. Some of the kinematic variables which can distinguish the signal events from those forming the background are shown in Figure \ref{fig:kins}. The event numbers for the signal and background processes are normalized to 1. $E_{T}^{{\rm Miss}}$ and $H_{T}$ represent the missing and hadronic transverse energies, respectively. $R_{{\rm ISR}}$ denotes the projection of missing transverse energy into the direction of jets emerging from initial state radiation (ISR). Finally $\Delta R_{{\rm min}}$ is the minimum of the angular distance between the variables denoted in the parentheses. The distribution for the total background is represented with the red curve, while the blue curve shows the distribution for the signal processes.

$E_{T}^{{\rm Miss}}$ is typically one of the key variables since the signal processes usually have more missing energy due to the presence of LSP neutralinos. However, this is true only when the mass difference between the stop and LSP neutralino is large enough, which is not applicable in our case. The mass difference in our benchmark point is about 200 GeV, and it does not yield a strong impact on the missing energy compared to the masses of the stop and LSP neutralino. As is seen in the panel plotting $E_{T}^{{\rm Miss}}$, the signal and background events lead to a peak at a similar scale, at which the missing energy is formed mostly by neutrinos. However, beyond this scale, the missing energy distribution exhibits a sharp decrease for the background events, while its change is relatively slow for the signal. It is observed because of the small but non-zero transverse momenta of the LSP neutralinos in the signal processes. Another variable can be formed by considering the ratio of the hadronic transverse energy to the missing energy as shown in the top-right panel. Even though this ratio yields similar peaks as observed in the missing energy distribution, it decreases much faster for the signal than for the background processes. 

Another important variable can arise from the ISR jets. In the compressed spectrum, most of the energy is consumed by the LSP neutralino mass which softens the top quark. In this case, the ISR jet and $E_{T}^{{\rm Miss}}$ are in the opposite directions. The projection of $E_{T}^{{\rm Miss}}$ onto the ISR jet momentum is expected to be maximum at about the ratio of the LSP neutralino mass to stop mass ($\sim 1$ in our case), while it is observed to be small and wider for the background processes \cite{ATLAS:2017drc}. This discussion for $R_{{\rm ISR}}$ can be observed in the bottom-left panel. Finally we also display the minimum angular distance between $E_{T}^{{\rm Miss}}$ and the jets in the final states. The distribution given in the bottom-right plane can provide an improvement to isolate the signal events by imposing a cut on the minimum angular distance between jets and $E_{T}^{{\rm Miss}}$, since the signal process exhibits a peak at a slightly lower value of $\Delta R_{{\rm min}}(j_{i},E_{T}^{{\rm Miss}})$.

\begin{table}[h!]
\centering
\setstretch{1.5}
\begin{tabular}{|c|c|c|c|c|} \hline
& $E_{T}^{{\rm Miss}}$ & $H_{T}/E_{T}^{{\rm Miss}}$ & $R_{{\rm ISR}}$ & $\Delta R_{{\rm min}}(j_{i},E_{T}^{{\rm Miss}})$ \\ \hline
Set 1 & $\geq$ 665.0 GeV & - & - & - \\ \hline
Set 2 & $\geq$ 665.0 GeV & $\leq$ 3.33 & - & - \\ \hline
Set 3 & $\geq$ 665.0 GeV & $\leq$ 3.33 & $\geq$ 0.64 & - \\ \hline
Set 4 & $\geq$ 665.0 GeV & $\leq$ 3.33 & $\geq$ 0.64 & $\leq$ 5.11 \\ \hline
\end{tabular}
\caption{The sets of cut flows imposed in our analyses. The cut on each variable is applied subsequently on top of the previous set.}
\label{tab:cuts}
\end{table}

Based on the results discussed with the plots in Figure \ref{fig:kins} we impose the cuts on the kinematic variables subsequently as listed in Table \ref{tab:cuts}. The final results are obtained after the fourth set of the cut flow, but we list them as different sets to discuss the effects from each cut in detail. Recall that these cuts are imposed after the pre-selection rules are applied. The calculated number of events for the FCC-hh experiments of 100 TeV COM are listed in Table \ref{tab:cutflow_fcc} after each set. The effectiveness of each cut set on the signal processes is parametrized by $\epsilon_{S}$, which is simply the ratio of the number of signal events as 

\begin{equation}
\epsilon_{S} \equiv \dfrac{{\rm Number ~ of ~ Events~(after~cut)}}{{\rm Number ~ of ~ Events~(pre\text{-}selection)}}~,
\label{eq:effS}
\end{equation}
and the significance $\mathcal{Z}$ is calculated through Eq.(\ref{eq:SS_0}). After including the relevant background processes enhanced with K-factors, the significance of the pre-selected signal events can be realized to be about 0.5 at about ten times larger integrated luminosity ($\mathcal{L}= 100$ fb$^{-1}$) than that reported in Table \ref{tab:benchs}.

\begin{table}[h!]
  \centering
  \setstretch{1.3}
  \resizebox{\textwidth}{!}{%
  \begin{tabular}{|lcccccccc|}
    \toprule
    Cut Flow & Signal &
    $t\bar{t}$ &
    $ttV$ &
    $tW$ &
    $VV$ &
    Total Bkg. &
    $\epsilon_S$ & $\mathcal{Z}$ \\
    \midrule
    Pre-selection &
    $2.87 \times 10^{4}$ &
    $3.58 \times 10^{9}$ &
    $7.18 \times 10^{6}$ &
    $3.28 \times 10^{8}$ &
    $1.48 \times 10^{8}$ &
    $4.06 \times 10^{9}$ &
    1.000 & 0.450 \\
    Set 1 &
    $5.00 \times 10^{3}$ &
    $4.69 \times 10^{5}$ &
    $1.95 \times 10^{4}$ &
    $6.90 \times 10^{5}$ &
    $6.38 \times 10^{4}$ &
    $1.24 \times 10^{6}$ &
    0.175 & 4.486 \\
    Set 2 &
    $4.72 \times 10^{3}$ &
    $3.52 \times 10^{5}$ &
    $1.54 \times 10^{4}$ &
    $4.45 \times 10^{5}$ &
    $5.32 \times 10^{4}$ &
    $8.65 \times 10^{5}$ &
    0.165 & 5.070 \\
    Set 3 &
    $4.60 \times 10^{3}$ &
    $3.52 \times 10^{5}$ &
    $1.45 \times 10^{4}$ &
    $4.00 \times 10^{5}$ &
    $4.26 \times 10^{4}$ &
    $8.09 \times 10^{5}$ &
    0.161 & 5.111 \\
    Set 4 &
    $4.60 \times 10^{3}$ &
    $3.52 \times 10^{5}$ &
    $1.43 \times 10^{4}$ &
    $3.56 \times 10^{5}$ &
    $4.26 \times 10^{4}$ &
    $7.64 \times 10^{5}$ &
    0.160 & 5.251 \\
    \bottomrule
  \end{tabular}%
  }
  \caption{The number of events for the signal and background processes realized at FCC-hh experiments of 100 TeV COM and $100$ fb$^{-1}$ integrated luminosity together with the effectiveness of cut flows and resultant significance for the signal processes.}
  \label{tab:cutflow_fcc}
\end{table}

\begin{figure}[h!]
\centering
\includegraphics[scale=0.5]{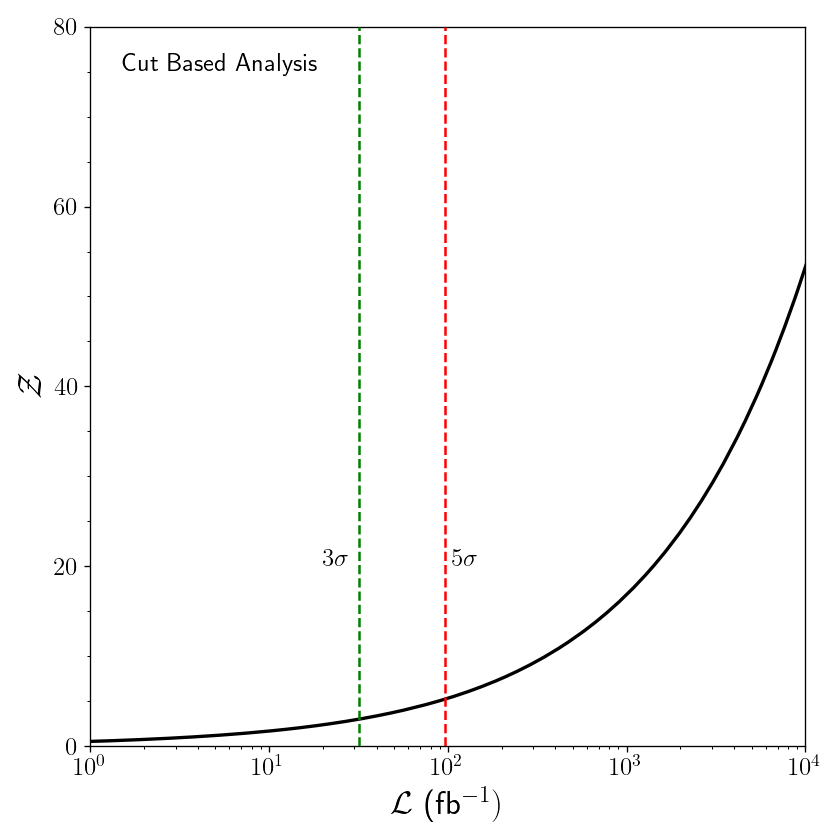}
\caption{The correlation between the significance and integrated luminosity ($\mathcal{L}$). The green dashed line indicates the luminosity value where the significance of the signal events can be realized within $3\sigma$, while the purple dashed line shows its probe within $5\sigma$ at the collider experiments proposed in FCC-hh.}
\label{fig:inlumcut}
\end{figure}

It can be seen easily from the results given in Table \ref{tab:cutflow_fcc} that the cut on the missing energy immediately increases the significance of the signal events by about 10 times, while the other sets can also provide minor improvements. Eventually, only about $16\%$ of the signal events can survive after the cuts in Set 4, which corresponds to $\mathcal{Z} \simeq 5.2$. Recall that we assume $\mathcal{L}= 100$ fb$^{-1}$ in calculation of significance. In this context, one can expect significantly improved results as the integrated luminosity increases, as shown in Figure \ref{fig:inlumcut}. According to the correlation between the significance and integrated luminosity, the compressed spectra of the stop and LSP neutralino can be probed up to about $3\sigma$ when $\mathcal{L}\simeq 32$ fb$^{-1}$, and $5\sigma$ probe can be realized for $\mathcal{L}\simeq 97$ fb$^{-1}$.

%\clearpage
\noindent{\bf BDT Improvement:}

Even though the cut-based analyses can yield a strong probe at the FCC-hh experiments, the results obtained above can be improved further through Boosted Decision Tree (BDT) analyses. We employ the BDT algorithm based on XGBoost (eXtreme Gradient Boosting) trained with $10^{5}$ events for each of the signal and background processes generated independently. In addition to the variables described above, the events are transferred to the BDT scan with $m_{CT}$ \cite{Tovey:2008ui}, $m_T\!\left(\ell,E_T^{\mathrm{miss}}\right)$ \cite{ATLAS:2024cmj}, $m_{T2}^W$ \cite{Bai:2012gs}, $m(b,l)$ and  $m_T\left(b_i,E_T^{\mathrm{miss}}\right)$ \cite{ATLAS:2017drc}. Note that the events are refined only by applying the pre-selection rules without any further cuts on the variables.

The expected impact from these variables can be summarized as follows: The number of events reaches its peak at about $ m_{CT} \approx \Delta m(\tilde{t}_{1},\tilde{\chi}_{1}^{0})$ for the signal, while its distribution is expected to be wider for the background. The transverse mass of the lepton and $E_{T}^{{\rm Miss}}$ ($m_T\!\left(\ell,E_T^{\mathrm{miss}}\right)$) cannot exceed the $W-$boson mass for the background, while it can take larger values for the signal processes, since the LSP neutralino contributes to the missing energy. A cut on $m_{T2}^W$ is utilized in such analyses to suppress the decay channels whose visible final state involves a single lepton. Since the presence of the ISR jet boosts the missing energy, it can provide a useful separation between the signal and background processes in the distribution over $m_T\left(b_i,E_T^{\mathrm{miss}}\right)$. 

In addition, one can consider the Razor variables which are useful when the processes involve invisible particles, especially when they have similar masses. Some of them can be defined as follows \cite{CMS:2016igb}:

\begin{equation}
\setstretch{1.7}
\begin{array}{rl}
M_{TR}^2 & = \dfrac{E_T^{\mathrm{Miss}}\left(p_{T1}+p_{T2}\right) - \vec{p}_T^{\,\mathrm{Miss}}\cdot\left(\vec{p}_{T1}+\vec{p}_{T2}\right)}{2} \\
M_R^2 & = \left(\left|\vec{p}_1\right| + \left|\vec{p}_2\right|\right)^2 - \left(p_{1z}+p_{2z}\right)^2 \\
R^2 &= \left(\dfrac{M_{TR}}{M_R}\right)^2
\end{array}
\label{eq:RMTR}
\end{equation}

If one assigns the top quarks to be megajets, whose transverse momenta are $p_{T1}$ and $p_{T2}$, $R^{2}$ could be useful in distinguishing the signal processes from the background. The boost on $E_{T}^{{\rm Miss}}$ from the ISR jet aligns it with these megajets, which maximizes  $M_{TR}^2$. In this case, one can realize $R^{2} \simeq R^{2}_{{\rm max}} = (m_{\tilde{t}_{1}}-m_{\tilde{\chi}_{1}^{0}})/m_{\tilde{\chi}_{1}^{0}}$. While most of the signal events are expected at $R^{2}\simeq R^{2}_{{\rm max}}$, the distribution of the background events over $R^{2}$ happens to be smoother. 

\begin{figure}[h!]
\centering
\includegraphics[scale=0.5]{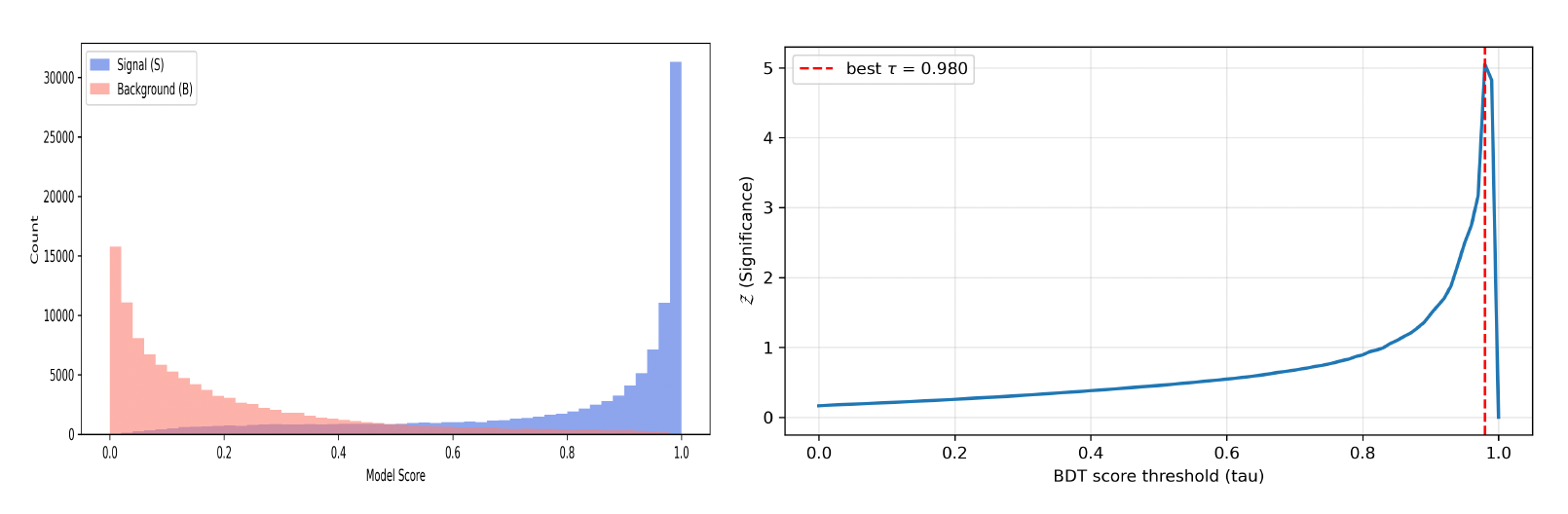}
\caption{The separation between the signal and background processes by BDT (left) and the best score threshold (right) obtained within our analyses.}
\label{fig:BDTscore}
\end{figure}

We transfer the pre-selected events to the BDT algorithm, which is trained together with the variables described above. We obtain the best models with the hyperparameters summarized  in Table \ref{tab:hypers}.

\begin{table}[htbp]
  \centering
  \label{tab:xgb_hyperparams}
  \begin{tabular}{lclc}
    \toprule
    Hyperparameter & Value & Hyperparameter & Value \\
    \midrule
    learning\_rate & $0.0408$ & gamma & $1.98$ \\
    max\_depth & $6$ & reg\_lambda & $0.431$ \\
    min\_child\_weight & $5.79$ & reg\_alpha & $3.28$ \\
    subsample & $0.752$ & ROC-AUC & 0.933 \\
    \bottomrule
  \end{tabular}
    \caption{The hyperparameters for the best XGBoost model.}
\label{tab:hypers}
\end{table}

\begin{figure}[h!]
\centering
\includegraphics[scale=0.5]{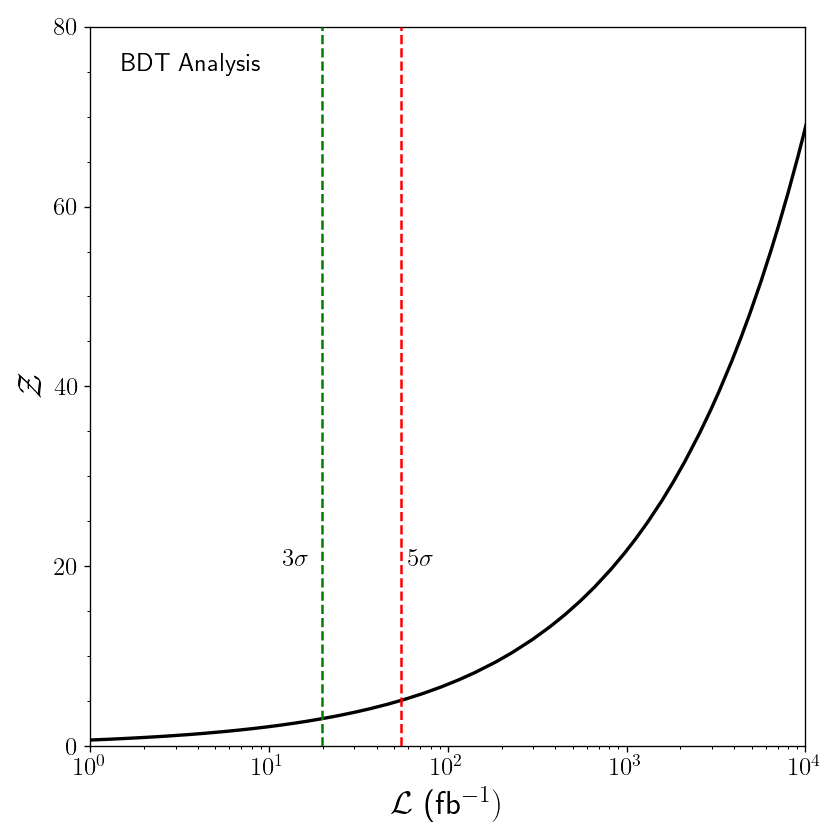}
\caption{The correlation between the significance and integrated luminosity ($\mathcal{L}$). The green dashed line indicates the luminosity value where the significance of the signal events can reach $3\sigma$, while the purple dashed line shows its probe within $5\sigma$ at the collider experiments proposed for FCC-hh.}
\label{fig:inlumcut2}
\end{figure}

After training, BDT can identify the signal processes over the background quite efficiently as shown in the left panel of Figure \ref{fig:BDTscore}. The best score in our analyses is obtained as $\tau = 0.98$, which is tested over the signal significance displayed in the right panel. With this optimal score in BDT, we realize a significant improvement in that the analyses can probe the signal up to about $3\sigma$ with an integrated luminosity of about $19.4$ fb$^{-1}$, and $5\sigma$ when $\mathcal{L} \simeq 54.4$ fb$^{-1}$. The significance of the signal in correlation with the integrated luminosity is given in Figure \ref{fig:inlumcut2}. 

{Before concluding, we should note that the analyses presented in this section were performed at the detector level using fast simulation. However, since the detector design for the FCC is still in progress, this study does not incorporate experimental systematic or detector-related uncertainties. Current studies on the model systematic uncertainties aim to incorporate pile-up effects \cite{Mangano:2022ukr} and the high-energy behavior of SM background processes \cite{Mangano:2016jyj}. Even though further refinements are needed, initial results indicate that in collisions at a 100 TeV center-of-mass (COM) energy, both pile-up effects and additional subleading background processes can yield a significant impact on the final sensitivity. In this context, our results rather reveal potential directions and prospects for new physics searches when the FCC experiment is launched. Our results can be recovered when the experimental searches reach the sensitivity in which the overall systematic uncertainties in the background modelling do not exceed $0.1\%$.}

\section{Conclusion}
\label{sec:conc}

We explore the implications within the high scale SUSY models based on the gauge symmetry $\PS$. If this gauge symmetry is broken into the SM gauge group near $\mgut$, one can maintain the gauge coupling unification approximately. In addition, these models also impose YU in their minimal construction. Previous studies have reported rather heavy mass spectra favored by YU. This results is also supported by the correct relic density constraint on the LSP neutralino measured by the Planck satellite. In our work, we include the NH terms which are mostly ignored in previous studies. These terms significantly change the profile of YU through the threshold corrections to the Yukawa couplings. In this context, we identify the YU solutions compatible with low fine-tuning, while the Higgsino-like LSP can be heavier because of the contributions to its mass from $\mu^{\prime}$. Even though this observation is interesting from naturalness point of view, the Higgsino-like LSP solutions receive strong strike from the DM observations in the standard DM scenarios.

The YU condition in the models of the \FTT~ class can be satisfied only when the gluino is lighter than about 1.2 TeV, which is more or less excluded by the current collider analyses. On the other hand, the NH contributions make the YU solutions possible in the regions with heavier gluinos, whose masses lie from about 2.2 to 10 TeV. The collider experiments with high luminosity are projected to probe these solutions up to about 2.5 TeV, while future experiments can improve the sensitivity up to about $m_{\tilde{g}} \lesssim 6$ TeV. The squarks of the first two generations are realized to be heavier than about 4 TeV, which are beyond the current sensitivity in the collider experiments.

The third-generation squarks, on the other hand, can be driven to lighter mass scales by the NH terms. Typical YU solutions usually lead to sbottoms with masses heavier than about 5 TeV, while it can be as light as about 1 TeV when the NH contributions are taken into account. The current analyses can exclude these solutions for $m_{\tilde{b}_{1}}\lesssim 1.5$ TeV. Similarly, the stop mass can also be as light as about 1.5 TeV. These solutions can be excluded up to about 1.2 TeV when the LSP neutralino is lighter than about 700 GeV. We also identify solutions in which the stop and LSP neutralino can be nearly degenerate at $m_{\tilde{t}_{1}}\simeq m_{\tilde{\chi}_{1}^{0}} \gtrsim 1.4$ TeV. Despite the fact that these solutions can escape from the common stop analyses, the recent analyses can improve the sensitivity up to about $m_{\tilde{t}_{1}}\lesssim 700$ GeV. 

In our analyses, we also require the relic density constraint to be satisfied, and we find the lightest possible stop to weigh about 1.7 TeV, with mass difference of about 200 GeV relative to the LSP neutralino. Such solutions are clearly beyond the reach of the current analyses, but they can be probed in future experiments such as those proposed for FCC-hh. We discuss possible probes for these solutions by considering the collider analyses projected for FCC-hh. In these analyses, we consider semi-leptonic final states emerging from stop pair-production. With suitable event selections based on cuts applied to the kinematic variables, we realize statistical significance of about $5\sigma$ when the integrated luminosity is increased to about $100$ fb$^{-1}$. We also perform a BDT analysis which results in such a large significance when the experimental analyses have lower luminosity as $\mathcal{L} \simeq 54$ fb$^{-1}$. {These results are rather preliminary, and they can be probed when the systematical uncertainties in background modeling are less than about $0.1\%$.}

\noindent{\bf Acknowledgment}

The work of BN and CSU is supported in part by the Scientific and Technological Research Council of Turkey (TUBITAK) Grant No. MFAG-125F122. The numerical calculations reported in this paper were partially performed at TUBITAK ULAKBIM, High Performance and Grid Computing Center (TRUBA resources).

\providecommand{\href}[2]{#2}\begingroup\raggedright\endgroup

%\clearpage
%\bibliographystyle{JHEP}
%\bibliography{NH}

\end{document}